\documentclass[12pt]{article}
\pdfoutput=1
\usepackage{amssymb}
\usepackage{epsfig}

\usepackage{graphicx}

\def\L{\mathcal L}
\def\e{\varepsilon}

\newcommand{\wt}{\widetilde}

\begin{document}

\def\a{\alpha}
\def\b{\beta}
\def\c{\chi}
\def\d{\delta}
\def\e{\epsilon}
\def\f{\phi}
\def\g{\gamma}
\def\h{\eta}
\def\i{\iota}
\def\j{\psi}
\def\k{\kappa}
\def\l{\lambda}
\def\m{\mu}
\def\n{\nu}
\def\o{\omega}
\def\p{\pi}
\def\q{\theta}
\def\r{\rho}
\def\s{\sigma}
\def\t{\tau}
\def\u{\upsilon}
\def\x{\xi}
\def\z{\zeta}
\def\D{\Delta}
\def\F{\Phi}
\def\G{\Gamma}
\def\J{\Psi}
\def\L{\Lambda}
\def\O{\Omega}
\def\P{\Pi}
\def\Q{\Theta}
\def\S{\Sigma}
\def\U{\Upsilon}
\def\X{\Xi}

\def\ve{\varepsilon}
\def\vf{\varphi}
\def\vr{\varrho}
\def\vs{\varsigma}
\def\vq{\vartheta}

\def\dg{\dagger}                                     
\def\ddg{\ddagger}                                   
\def\wt#1{\widetilde{#1}}                    
\def\mt{\widetilde{m}_1}
\def\mti{\widetilde{m}_i}
\def\rt{\widetilde{r}_1}
\def\mtt{\widetilde{m}_2}
\def\mttt{\widetilde{m}_3}
\def\rtt{\widetilde{r}_2}
\def\mb{\overline{m}}
\def\VEV#1{\left\langle #1\right\rangle}        
\def\be{\begin{equation}}
\def\ee{\end{equation}}
\def\ds{\displaystyle}
\def\ra{\rightarrow}

\def\bea{\begin{eqnarray}}
\def\eea{\end{eqnarray}}
\def\NO{\nonumber}
\def\Bar#1{\overline{#1}}


\def\pl#1#2#3{Phys.~Lett.~{\bf B {#1}} ({#2}) #3}
\def\np#1#2#3{Nucl.~Phys.~{\bf B {#1}} ({#2}) #3}
\def\prl#1#2#3{Phys.~Rev.~Lett.~{\bf #1} ({#2}) #3}
\def\pr#1#2#3{Phys.~Rev.~{\bf D {#1}} ({#2}) #3}
\def\zp#1#2#3{Z.~Phys.~{\bf C {#1}} ({#2}) #3}
\def\cqg#1#2#3{Class.~and Quantum Grav.~{\bf {#1}} ({#2}) #3}
\def\cmp#1#2#3{Commun.~Math.~Phys.~{\bf {#1}} ({#2}) #3}
\def\jmp#1#2#3{J.~Math.~Phys.~{\bf {#1}} ({#2}) #3}
\def\ap#1#2#3{Ann.~of Phys.~{\bf {#1}} ({#2}) #3}
\def\prep#1#2#3{Phys.~Rep.~{\bf {#1}C} ({#2}) #3}
\def\ptp#1#2#3{Progr.~Theor.~Phys.~{\bf {#1}} ({#2}) #3}
\def\ijmp#1#2#3{Int.~J.~Mod.~Phys.~{\bf A {#1}} ({#2}) #3}
\def\mpl#1#2#3{Mod.~Phys.~Lett.~{\bf A {#1}} ({#2}) #3}
\def\nc#1#2#3{Nuovo Cim.~{\bf {#1}} ({#2}) #3}
\def\ibid#1#2#3{{\it ibid.}~{\bf {#1}} ({#2}) #3}

\title{
\vspace*{-1mm}
\bf A full analytic solution of $SO(10)$-inspired leptogenesis}
\author{
{\Large Pasquale Di Bari$^1$ and Michele Re Fiorentin$^{1,2}$}
\\
$^1$ {\it\small Physics and Astronomy}, 
{\it\small University of Southampton,} 
{\it\small  Southampton, SO17 1BJ, U.K.} \\
$^2$ {\it\small Center for Sustainable Future Technologies}, \\
{\it\small Istituto Italiano di Tecnologia, corso Trento 21, 10129 Torino, Italy}
}

\maketitle \thispagestyle{empty}

\vspace{-10mm}

\begin{abstract}
Recent encouraging experimental results on neutrino mixing parameters prompt further
investigation on $SO(10)$-inspired leptogenesis  and on the associated strong thermal solution
that has correctly predicted a non-vanishing reactor mixing angle,
it further predicts $\sin\d \lesssim 0$,  now supported by recent results at $\sim 95\%$ C.L., 
normally ordered neutrino masses and atmospheric mixing angle in the first octant, best fit 
results in latest global analyses. 
Extending a recent analytical procedure, we account for the mismatch between the Yukawa
basis and the weak basis, that in $SO(10)$-inspired models is described by a CKM-like 
unitary transformation $V_L$, obtaining a full analytical solution that  provides useful  insight 
and reproduces accurately all numerical results\footnote{In this latest version we  
corrected a simple typo present on the published version: the right-hand sides of Eqs.~(47) and (48) were erroneously swapped. Of course in the code that generated panels of Fig.~2 the correct expressions were used, otherwise the perfect agreement between analytical and numerical plots would not have been obtained.
We wish to thank Rome Samanta for having spotted the typo.}, paving the way for future inclusion of different sources of theoretical uncertainties and for a statistical analysis of the constraints. 
We show how  muon-dominated solutions  appear for large values of the lightest
neutrino mass in the range $(0.01$--$1)\,{\rm eV}$ but also how they necessarily require a mild fine tuning
in the seesaw relation.   For the dominant (and untuned) tauon-dominated
solutions  we show analytically how, turning on $V_L \simeq V_{CKM}$, some of the constraints on the low energy neutrino parameters  get significantly relaxed. In particular we show how the upper bound on the atmospheric neutrino mixing angle in the strong thermal solution gets relaxed from $\theta_{23} \lesssim 41^{\circ}$ to $\theta_{23} \lesssim 44^{\circ}$, an important effect  in the light
of the most recent NO$\nu$A, T2K and IceCube results. 
\end{abstract}

\newpage
\section{Introduction}

The latest results from the LHC  show no evidence for new physics at the TeV scale.
Analogously, negative results also come from direct dark matter, cLFV and  electric dipole moments searches. 
Potential manifestations of new physics are given by the long standing muon $g-2$ anomaly and by the recent anomalies 
reported in $B$ decays \cite{ICHEP} but  
more solid evidence is required and they are indeed currently regarded as anomalies.
On the other hand robust motivations for extending the Standard Model come from  neutrino masses and mixing and   
from the cosmological puzzles. In the absence of new physics 
at the TeV scale or below, it is reasonable to think that their solution is related to the existence 
of higher energy scales.   In particular  a combined explanation of neutrino masses and mixing,  
from a conventional high energy type I seesaw mechanism \cite{seesaw},
and of the matter-antimatter asymmetry of the Universe
from (consequentially high energy scale) leptogenesis \cite{fy},
should be currently regarded as the simplest and attractive  possibility. 

Interestingly,  latest neutrino oscillation experiments global analyses 
also  seem to support $C\!P$ violation in left-handed (LH) neutrino mixing 
(at $95\%\, {\rm C.L.}$ in \cite{marrone} and at $70\% \, {\rm C.L.}$ in \cite{nufit}).
Though this is not a sufficient condition for the existence of a source of $C\!P$
violation for successful leptogenesis, if confirmed,
it would be still an important result since it would make quite plausible
the presence of $C\!P$ violation also in heavy right-handed (RH) neutrino mixing, 
the natural dominant source of $C\!P$
violation for leptogenesis (barring special scenarios). 
\footnote{Conversely, $C\!P$  conservation in LH neutrino 
mixing   would legitimately cast some doubts on it.}
In addition, the exclusion of quasi-degenerate  light neutrino masses
can be also regarded as a positive experimental outcome for minimal scenarios of leptogenesis, 
based on type I seesaw mechanism and thermal RH neutrino production, 
since the bulk of solutions  requires values of neutrino masses $m_i \lesssim {\cal O}(0.1)\,{\rm eV}$ \cite{upperbound,bounds}, 
even when charged lepton \cite{flavour} and heavy neutrino \cite{geometry} flavour effects are taken into account.
\footnote{As we will discuss in detail, muon-dominated solutions in $SO(10)$-inspired leptogenesis are found for
$m_i$ as large as $\sim 1\,{\rm eV}$, but these solutions suffer of some fine-tuning, as we will notice,
and  are certainly less interesting than  tauon-dominated solutions representing
 the bulk of solutions  and respecting the upper bound $m_i \lesssim {\cal O}(0.1\,{\rm eV})$.} 
Therefore, this current phenomenological picture certainly encourages further investigation on high energy 
scale scenarios of leptogenesis.  

On the other hand the possibility to test more stringently leptogenesis and even have any hope to prove it, 
seems necessarily to rely on the identification of specific scenarios, 
possibly emerging from well motivated theoretical frameworks. 
This is in order to reduce the number of independent parameters, increasing the predictive power 
and over-constraining the seesaw parameter
space. The sharper the predictions are, the lower the probability  that these
are just a mere coincidence.   
This challenging strategy has been strongly boosted by the measurement of a non-vanishing value of the reactor mixing angle, sufficiently large to make possible the measurements of the 
unknown parameters in the leptonic mixing matrix:  $C\!P$ violating Dirac phase, 
neutrino mass ordering and a determination of the deviation of the atmospheric mixing angle 
from the maximal value. 

The latest results from the NO$\nu$A long baseline 
experiment favour a  $\sim 5^{\circ}$ deviation 
of the atmospheric mixing angle from maximal
mixing \cite{NOvA}, while the results from the  T2K long baseline experiment \cite{T2K}
and from the IceCube neutrino detector \cite{IceCube} 
do not find evidence of such deviation  so far, so that a mild tension exists but 
still within  $90\%\,{\rm C.L.}$
At the same time both experiments strengthen 
the support for negative values of $\sin\d$. 
Moreover they also show an emergence for a slight preference for normally ordered neutrino masses. When all results are combined, 
two recent global analyses 
find that first octant for atmospheric  mixing angle with normally ordered neutrino
masses (NO) emerges as a best fit solution, though the preference over inverted
ordered neutrino masses (IO), allowing both first and second octant, 
is currently slight, at the level of $\sim 1.7\,\s$ \cite{marrone} or even less \cite{nufit}.

Intriguingly, this emerging potential experimental set of results for the unknown neutrino oscillation parameters 
nicely supports the expectations from the so 
called strong thermal $SO(10)$-inspired leptogenesis solution \cite{strongSO10},  
indeed strictly requiring NO,  atmospheric mixing angle in the first octant and favouring 
negative values of $\sin\d$ for sufficiently large values of the atmospheric mixing angle.\footnote{The solution also requires non-vanishing $\theta_{13}$ for large values of the initial pre-existing
asymmetry $N^{\rm p,i}_{B-L} \gtrsim 0.001$, a result preliminarily  presented in \cite{DESY} before
the discovery from nuclear reactors. }

This solution relies on two   independent conditions and it is highly non trivial that they can be satisfied simultaneously. 
The first one, on the model building side, is the {\em $SO(10)$-inspired condition} \cite{SO10inspired}, and
it corresponds to assume that the  Dirac neutrino mass matrix is {\em not too different} from the up-quark mass
matrix, a typical feature of different grand-unified models such as $SO(10)$ models.\footnote{As we will see in more detail, even without imposing the strong thermal
condition, the $SO(10)$-inspired condition already strongly favours NO and to less extent the atmospheric
mixing angle in the first octant.}
The second assumption is dictated purely  by a cosmological
requirement, the independence of the final asymmetry of the initial conditions (the {\em strong thermal leptogenesis condition}).  
The latter, in the case of hierarchical RH neutrino mass patterns, 
is satisfied only for  quite a specific case, the tauon $N_2$-dominated scenario \cite{problem} and, as we said, 
it is highly not trivial that this is realised within $SO(10)$-inspired models. 
 If future data will confirm  NO together with
$\sin\d < 0$ and atmospheric mixing angle in the first octant, 
the statistical significance of the agreement between theoretical predictions 
and experimental results would be very interesting, since
the probability to find by chance such an agreement with the
strong thermal $SO(10)$-inspired leptogenesis solution is lower than $\sim 5\%$ \cite{strongSO10}. 


A full analytical description of $SO(10)$-inspired leptogenesis is greatly helpful in different respects.
First, it provides a useful analytical insight  able to clarify different interesting  aspects of 
$SO(10)$-inspired leptogenesis, as we will discuss in detail. On more practical grounds,
the expected future improvement in the  determination of the neutrino mixing parameters
clearly calls for an analogous improvement in the theoretical predictions with a reduction of the
theoretical uncertainties.  To this extent, for an inclusion of more subtle effects in the derivation of the
low energy neutrino constraints, a full analytical calculation of the final asymmetry  
allows a fast generation of solutions, something essential also for a precise
statistical derivation of the constraints, so far qualitatively derived just from scatter plots. 
Driven by these motivations, in this paper we extend the analytic procedure of \cite{SO10decription},
taking into account the mismatch between the Yukawa basis and the weak basis. This will allow to reproduce
with great accuracy  all results obtained only numerically so far.\footnote{In the paper we will consider
a non-supersymmetric framework. For a  detailed discussion on the 
supersymmetric extension we refer the reader to
\cite{susy}, where it has been shown that constraints get significantly relaxed only at large
values $\tan\b \gtrsim 15$. The analytical results that we discuss here can be easily exported
to the supersymmetric case.}
The paper is organised as follows. 
In Section 2 we review the seesaw type I mechanism and current neutrino oscillation data. 
In Section 3 we discuss $SO(10)$-inspired leptogenesis extending the analytical procedure discussed
in \cite{SO10decription} taking into account the  mismatch, described by a unitary matrix $V_L$, between the Yukawa basis, where the neutrino Dirac mass matrix is diagonal, and the weak basis, where the charged lepton mass matrix is diagonal.\footnote{We summarise in the Appendix the set of expressions that allow a full general analytical calculation of the asymmetry in $SO(10)$-inspired leptogenesis.}
In this way we obtain some general results that in Section 4 we specialise to reproduce a few different effects governed by the matrix $V_L$ including the application to strong thermal
leptogenesis showing how the upper bound on the atmospheric mixing angle gets relaxed.
Finally in Section 5 we draw the conclusions. 

\section{Seesaw and low energy neutrino parameters}

Augmenting the SM with three RH neutrinos $N_{i R}$
with Yukawa couplings $h$ and a Majorana mass term M, in the flavour basis, where both
charged lepton mass matrix $m_{\ell}$ and $M$ are diagonal, one can write   the leptonic mass terms 
generated after spontaneous symmetry breaking as ($\a=e,\m,\t$ and $i=1,2,3$)
\be
- {\cal L}_{M} = \,  \overline{\a_L} \, D_{m_{\ell}}\,\a_R + 
                              \overline{\nu_{\a L}}\,m_{D\a i} \, N_{i R} +
                               {1\over 2} \, \overline{N^{c}_{i R}} \, D_{M} \, N_{i R}  + \mbox{\rm h.c.}\, ,
\ee
where $D_{m_{\ell}} \equiv {\rm diag}(m_e,m_{\m},m_{\t})$, $D_{M}\equiv {\rm diag}(M_1,M_2,M_3)$
and $m_{D}$ is the neutrino Dirac mass matrix. 
In the seesaw limit, $M \gg m_D$, the mass spectrum splits into two sets of Majorana eigenstates:
a light set with masses $m_{1} \leq m_{2} \leq m_3$ given by the seesaw formula \cite{seesaw}
\be\label{seesaw}
D_m =  U^{\dagger} \, m_D \, {1\over D_M} \, m_D^T  \, U^{\star}  \,   ,
\ee
with $D_m = {\rm diag}(m_1,m_2,m_3)$, and a heavy set with masses basically
coinciding with $D_M$. The matrix $U$, which diagonalises the light neutrino mass
matrix $m_{\nu} = -m_D\,M^{-1}\,m_D^T$ in the weak basis, can then be identified
with the PMNS lepton mixing matrix. 
For NO, this can be parameterised
in terms of the usual mixing angles $\theta_{ij}$, the Dirac phase $\d$
and the Majorana phases $\rho$ and $\s$, as
\be
U=  \left( \begin{array}{ccc}
c_{12}\,c_{13} & s_{12}\,c_{13} & s_{13}\,e^{-{\rm i}\,\d} \\
-s_{12}\,c_{23}-c_{12}\,s_{23}\,s_{13}\,e^{{\rm i}\,\d} &
c_{12}\,c_{23}-s_{12}\,s_{23}\,s_{13}\,e^{{\rm i}\,\d} & s_{23}\,c_{13} \\
s_{12}\,s_{23}-c_{12}\,c_{23}\,s_{13}\,e^{{\rm i}\,\d}
& -c_{12}\,s_{23}-s_{12}\,c_{23}\,s_{13}\,e^{{\rm i}\,\d}  &
c_{23}\,c_{13}
\end{array}\right)
\, {\rm diag}\left(e^{i\,\rho}, 1, e^{i\,\sigma}
\right)\,   .
\ee
For IO, since we are defining $m_1 \leq m_2 \leq m_3$,
this should be replaced by the column permuted matrix
\be
U^{\rm (IO)} = 
U^{\rm (NO)} \, 
\left(\begin{array}{ccc}
0 & 1 & 0\\
0 & 0 & 1\\
1 & 0 & 0  
\end{array}\right) \,   .
\ee
However, in this paper we will focus on the NO case, since IO
is only marginally allowed 
imposing just successful $SO(10)$-inspired leptogenesis \cite{riotto2}
\footnote{It is allowed only at quite large values of $m_1 \gtrsim 10^{-2+0.14\,(52-\theta_{23}/^{\circ})}\,{\rm meV}$, so that for example using a more aggressive upper bound from the same
{\em Planck} collaboration $\sum_i \, m_i \lesssim 0.17\,{\rm eV}$ \cite{planck}, translating into 
$m_1 \lesssim 0.04\,{\rm eV}$ for IO, the allowed region is almost completely ruled out.}
and it is completely excluded imposing in addition the strong thermal leptogenesis condition.
In the case of NO, latest neutrino oscillation experiments global analyses find for the mixing angles and the 
leptonic Dirac phase $\d$,  the following best fit values, $1\s$ errors and $3\s$ intervals \cite{marrone}: 
\bea\label{expranges}
\theta_{13} & = &  8.45^{\circ}\pm 0.15^{\circ} \, \;\;  [8.0^{\circ}, 9.0^{\circ}] \,  , \\ \nonumber
\theta_{12} & = &  33^{\circ}\pm 1^{\circ} \,  \;\;  [30^{\circ}, 36^{\circ}]  \,  , \\ \nonumber
\theta_{23} & = &  {41^{\circ}} \pm {1^{\circ}} \,  \;\;  [38^{\circ}, 51.65^{\circ}]  \,  ,  \\ \nonumber
\d & = &  {-0.62 \pi \pm 0.2\pi} \,  \;\;  [-1.24\pi, 0.17\pi] \,  .
\eea 
It is interesting that there is already an excluded interval, $\d\ni [0.17\,\pi, 0.76\pi]$ at $3\,\s$,
and that  $\sin\d > 0$ is excluded at  $2\s$  favouring  $\sin\d < 0$
(in \cite{nufit} a lower statistical significance is found).  Of course there are no 
experimental constraints on the Majorana phases. 
Neutrino oscillation experiments also measure two  mass squared differences, finding 
for the solar neutrino mass scale $m_{\rm sol}\equiv \sqrt{m^{\, 2}_2 - m_1^{\, 2}} = (8.6\pm 0.1)\,{\rm meV}$
and for the atmospheric neutrino mass scale $m_{\rm atm}\equiv \sqrt{m^{\, 2}_3 - m_2^{\, 2}} = (49.5\pm 0.05)\,{\rm meV}$ \cite{marrone}.

There is no signal from neutrinoless double beta ($0\nu\b\b$) decay experiments that, therefore, 
place an upper bound on the effective $0\nu\b\b$ neutrino mass defined as
\be
m_{ee} \equiv |m_{\nu ee}| = |U_{e1}^2 \,m_1 + U_{e2}^2 \,m_2 + U_{e3}^2 \,m_3| \,   .
\ee 
Currently, the most stringent reported upper bound comes from the KamLAND-Zen collaboration finding, at
$90\%\,{\rm C.L.}$, 
$m_{ee} \leq (61 \mbox{--} 165)\,{\rm meV}$  \cite{kamlandzen},
where the range accounts for nuclear matrix element uncertainties. 

Cosmological observations place an upper bound on the 
sum of the neutrino masses. The {\em Planck} satellite collaboration 
obtains a robust stringent upper bound $\sum_i m_i \lesssim 230\,{\rm meV}$ at $95\% {\rm C.L.}$
\cite{planck} that, taking into account neutrino oscillation experimental determination
of the solar and atmospheric neutrino mass scales, translates into an upper bound on the 
lightest neutrino mass $m_1 \lesssim 70\,{\rm meV}$. 

\section{$SO(10)$-inspired leptogenesis}

The neutrino Dirac mass matrix can be diagonalised (singular value decomposition) as
\be\label{svd}
m_D = V^{\dagger}_L \, D_{m_D} \, U_R \,  ,
\ee
where $D_{m_D} \equiv {\rm diag}(m_{D1},m_{D2},m_{D3})$ and where
$V_L$ and $U_R$ are the two unitary matrices transforming respectively the
LH and RH neutrino fields from the 
flavour basis (where $m_{\ell}$ and $M$ are diagonal) 
to the Yukawa basis (where $m_D$ is diagonal).  

If we parameterise the neutrino Dirac masses $m_{Di}$ in terms of the up quark masses, 
\be\label{alphai}
(m_{D 1}, m_{D 2}, m_{D 3})=(\a_1 \, m_{\rm u}, \a_2\, m_{\rm c}, \a_3 \, m_{\rm t})  \,  ,
\ee
we can impose so called {\em $SO(10)$-inspired conditions} defined as:
\begin{itemize}
\item[-] $m_{D3} \gg m_{D2} \gg m_{D_1}$, implying $\a_i = {\cal O}(0.1$--$10)$ \,  ,
\item[-] $I\leq V_L \lesssim V_{CKM}$   .
\end{itemize}
The latter should be read in a way that  parameterising $V_L$
in the same way as the leptonic mixing matrix $U$, the three
mixing angles $\theta_{12}^L$, $\theta_{23}^L$ and $\theta_{13}^L$
cannot have values  much larger than the three mixing angles
in the CKM matrix.
\footnote{Precisely we adopt: $\theta_{12}^L \leq  13^{\circ} \simeq \theta_{12}^{CKM} \equiv \theta_c$, 
$\theta_{23}^L \leq  2.4^{\circ}\simeq \theta_{23}^{CKM}$, $\theta_{13}^L \leq  0.2^{\circ}\simeq \theta_{13}^{CKM}$. However, notice that the validity of our analytical solution goes beyond these ranges of values for the 
mixing angles in the $V_L$. We will discuss this point in greater detail in the Appendix.}

Inserting the singular value decomposed form for $m_D$ Eq.~(\ref{svd}) into the
seesaw formula Eq.~(\ref{seesaw}), one obtains
\be\label{invM}
M^{-1} \equiv  U_R \, D_M \, U_R^T = - D_{m_D}^{-1} \, \widetilde{m}_{\nu} \, D_{m_D}^{-1} \,  ,
\ee
where $M \equiv U^{\star}_R\,D_M\,U^{\dagger}_R$ and 
$\widetilde{m}_{\nu} \equiv V_L\,m_{\nu}\,V_L^T$ 
are respectively  the Majorana mass matrix and  the light neutrino mass matrix 
in the Yukawa basis.  Diagonalising the matrix on the RH side of Eq.~(\ref{invM}),
one can express the RH neutrino masses and the RH neutrino mixing matrix $U_R$
in terms of $m_{\nu}$, $V_L$ and the three $\a_i$. 

The analytical procedure discussed in \cite{SO10decription}, 
within the approximation $V_L \simeq I$, gets easily generalised for $V_L \neq I$
replacing $m_{\nu}\rightarrow \widetilde{m}_{\nu}$ \cite{afs} and in this case 
one finds for the three RH neutrino masses 
\footnote{As pointed out in \cite{SO10decription}, the validity of these results 
relies on hierarchical RH neutrino masses, $M_3 \gg M_2 \gg M_1$ and breaks down in the close
vicinity of crossing level solutions found in \cite{afs} where either 
$|\widetilde{m}_{\nu 11}|$ or $|(\widetilde{m}_{\nu}^{-1})_{33}|$ or both vanish. However, 
as we will point out, 
 when $|\widetilde{m}_{\nu 11}|$ or $|(\widetilde{m}_{\nu}^{-1})_{33}|$ vanish separately, corresponding to
 $M_1 \simeq M_2$ and $M_2 \simeq M_3$ respectively, successful leptogenesis is not attained, and the case when they both get very small, leading to a compact spectrum $M_1 \sim M_2 \sim M_3$, 
necessarily implies a huge fine-tuning in the seesaw formula since in this case the orthogonal matrix elements become huge, as we are going to show. For this reason a hierarchical spectrum condition is not restrictive at all. We will be back on this point.}
\be\label{Mi}
M_1    \simeq   {m^2_{D1} \over |\widetilde{m}_{\nu 11}|} \, , \;\;
M_2  \simeq    {m^2_{D2} \over m_1 \, m_2 \, m_3 } \, {|\widetilde{m}_{\nu 11}| \over |(\widetilde{m}_{\nu}^{-1})_{33}|  } \,  ,  \;\;
M_3  \simeq   m^2_{D3}\,|(\widetilde{m}_{\nu}^{-1})_{33}|  ,
\ee
and for the RH neutrino mixing matrix
\be\label{UR}
U_R \simeq  
\left( \begin{array}{ccc}
1 & -{m_{D1}\over m_{D2}} \,  {\widetilde{m}^\star_{\nu 1 2 }\over \widetilde{m}^\star_{\nu 11}}  & 
{m_{D1}\over m_{D3}}\,
{ (\widetilde{m}_{\n}^{-1})^{\star}_{13}\over (\widetilde{m}_{\n}^{-1})^{\star}_{33} }   \\
{m_{D1}\over m_{D2}} \,  {\widetilde{m}_{\nu 12}\over \widetilde{m}_{\nu 11}} & 1 & 
{m_{D2}\over m_{D3}}\, 
{(\widetilde{m}_{\n}^{-1})_{23}^{\star} \over (\widetilde{m}_{\n}^{-1})_{33}^{\star}}  \\
 {m_{D1}\over m_{D3}}\,{\widetilde{m}_{\nu 13}\over \widetilde{m}_{\nu 11}}  & 
- {m_{D2}\over m_{D3}}\, 
 {(\widetilde{m}_\nu^{-1})_{23}\over (\widetilde{m}_\nu^{-1})_{33}} 
  & 1 
\end{array}\right) 
\,  D_{\Phi} \,  ,
\ee
where the three phases in 
$D_{\phi} \equiv {\rm diag}(e^{-i \, {\Phi_1 \over 2}}, e^{-i{\Phi_2 \over 2}}, e^{-i{\Phi_3 \over 2}})$ 
are given by
\be
\Phi_1 = {\rm Arg}[-\widetilde{m}_{\nu 11}^{\star}] \,  , \; \;
\Phi_2 = {\rm Arg}\left[{\widetilde{m}_{\nu 11}\over (\widetilde{m}_{\nu}^{-1})_{33}}\right] -2\,(\rho+\s)-2\,(\rho_L + \s_L) \, , \;  \;
\Phi_3 = {\rm Arg}[-(\widetilde{m}_{\nu}^{-1})_{33}] \,  .
\ee
It should be noticed how the Majorana phases $\rho_L$ and $\s_L$ 
enter directly the expression for the RH neutrino Majorana phases (more precisely in $\Phi_2$) independently of the
values of the mixing angles $\theta_{ij}^L$. 
It will prove convenient to introduce a matrix
\be\label{URA}
A \equiv  
\left( \begin{array}{ccc}
1 & - \,  {\widetilde{m}^\star_{\nu 1 2 }\over \widetilde{m}^\star_{\nu 11}}  & 
{ (\widetilde{m}_{\n}^{-1})^{\star}_{13}\over (\widetilde{m}_{\n}^{-1})^{\star}_{33} }   \\
 {\widetilde{m}_{\nu 12}\over \widetilde{m}_{\nu 11}} & 1 &  
{(\widetilde{m}_{\n}^{-1})_{23}^{\star} \over (\widetilde{m}_{\n}^{-1})_{33}^{\star}}  \\
{\widetilde{m}_{\nu 13}\over \widetilde{m}_{\nu 11}}  & 
-  {(\widetilde{m}_\nu^{-1})_{23}\over (\widetilde{m}_\nu^{-1})_{33}} 
  & 1 
\end{array}\right) 
\,  D_{\Phi} \,    ,
\ee
such that the elements of $U_R$ can be written in the form 
\be
U_{Rij} =  {{\rm min}[m_{Di}, m_{Dj}] \over {\rm max}[m_{Di}, m_{Dj}]}\,A_{ i j}   \,   .
\ee

One can also derive an  expression for the 
orthogonal matrix starting from its definition
$\O = D_m^{-{1\over 2}}\, U^{\dagger} \, m_D \,  D_M^{-{1\over 2}} $ \cite{casas}
that, using Eq.~(\ref{svd}), becomes \cite{riotto1}
\be
\O= D_m^{-{1\over 2}}\, U^{\dagger} \, V_L^{\dagger} \, D_{m_D} \, U_R \, D_M^{-{1\over 2}} \,  .
\ee
In terms of matrix elements this can be written as 
\be\label{Oij}
\O_{ij} \simeq {1\over \sqrt{m_i \, M_j}} \, 
\sum_k \, m_{D l} \, U^{\star}_{ki}\,V^{\star}_{L\,lk}\,U_{R \, k j} \,  ,
\ee
from which one finds 
\footnote{This improves the analytical expression given in \cite{A2Z} 
where the approximation $W \simeq U$ was used. We checked that this
analytic expression perfectly reproduces the numerical results. This expression shows explicitly
how approaching the crossing level solutions, for vanishing 
$|\widetilde{m}_{\nu 11}|$ or $|(\widetilde{m}_{\nu}^{-1})_{33}|$, 
the $|\O_{ij}^2|$'s become huge and this corresponds to very fine-tuned cancellations in the seesaw formula.}
\be\label{Omegaapp}
\hspace{-9mm}\O \simeq 
\left( \begin{array}{ccc}
{(\widetilde{m}_{\nu}\,W^{\star})_{11} \over \sqrt{- m_1 \, \widetilde{m}_{\nu 11}}} & 
\sqrt{m_2\,m_3\,(\widetilde{m}_{\nu}^{-1})_{33} \over \widetilde{m}_{\nu 11}}\,
\left(W^{\star}_{2 1} - W^{\star}_{31}\,{{(\widetilde{m}_{\nu}^{-1})_{23}}\over (\widetilde{m}_{\nu}^{-1})_{33}}\right) \, e^{i\,(\rho+\s +\r_L + \s_L)} & 
{W^{\star}_{31}\over \sqrt{m_1\,(\widetilde{m}_{\nu}^{-1})_{33}}} \\
 {(\widetilde{m}_{\nu}\,W^{\star})_{12} \over \sqrt{- m_2 \, \widetilde{m}_{\nu 11}}} & 
\sqrt{m_1\,m_3\,(\widetilde{m}_{\nu}^{-1})_{33} \over \widetilde{m}_{\nu 11}}\,
\left(W^{\star}_{22} - W^{\star}_{32}\,{{(\widetilde{m}_{\nu}^{-1})_{23}}\over (\widetilde{m}_{\nu}^{-1})_{33}}\right)  \, e^{i\,(\rho+\s +\r_L + \s_L)}
& {W^{\star}_{32}\over \sqrt{m_2\,(\widetilde{m}_{\nu}^{-1})_{33}}}  \\
 {(\widetilde{m}_{\nu}\,W^{\star})_{13} \over \sqrt{- m_3 \, \widetilde{m}_{\nu 11}}}  & 
\sqrt{m_1\,m_2\,(\widetilde{m}_{\nu}^{-1})_{33} \over \widetilde{m}_{\nu 11}}\,
\left(W^{\star}_{2 3} - W^{\star}_{3 3}\,{{(\widetilde{m}_{\nu}^{-1})_{23}}\over (\widetilde{m}_{\nu}^{-1})_{33}}\right)  \, e^{i\,(\rho+\s +\r_L + \s_L)}
& {W^{\star}_{33}\over \sqrt{m_3\,(\widetilde{m}_{\nu}^{-1})_{33}}}  
\end{array}\right)  \,   ,
\ee
where we defined $W \equiv V_L \, U$.

Let us now discuss the calculation of the asymmetry.  Since in Section 4 we will also be interested in those
solutions satisfying, in addition to successful leptogenesis, also the strong thermal condition,
we can write the  final asymmetry  as the sum of two terms,
\be\label{2terms}
N_{B-L}^{\rm f} = N_{B-L}^{\rm p,f} + N_{\rm B-L}^{\rm lep,f}  \,  .
\ee
The first term is the relic value of the pre-existing asymmetry,
the second is the asymmetry generated from leptogenesis. 
This  of course would translate into a baryon-to-photon number ratio
also given by the sum of two contributions, $\eta_B^{\rm p}$ and $\eta_B^{\rm lep}$
respectively. The typical assumption 
is that the initial pre-existing asymmetry, after inflation and prior to leptogenesis, is negligible. 
We also consider the possibility that some external mechanism might have generated
a large value of the  {\em initial} pre-existing asymmetry, $N_{B-L}^{\rm p,i}$,
between the end of inflation and the onset of leptogenesis, 
i.e. a  value that would translate, in the absence of any wash-out, into 
a sizeable value of $\eta_B^{\rm p}$.
The strong thermal leptogenesis condition requires  that
this initial value of the pre-existing asymmetry is efficiently washed out by RH
neutrinos wash-out processes in a way that the final value of $\eta_B$ is
dominated by $\eta_B^{\rm lep}$. 
\footnote{For definiteness we adopt a criterium $\eta_B^{\rm p} < 0.1\,\eta_B^{\rm lep}$ but
in any case  the
constraints on low energy neutrino parameters depend only logarithmically on the  precise
maximum allowed value for $\eta_B^{\rm p}/\eta_B^{\rm lep}$.}
The predicted value of the baryon-to-photon number ratio is then entirely explained
by the contribution from leptogenesis,
\be
\eta_B^{\rm lep} =a_{\rm sph}\,{N_{B-L}^{\rm lep,f}\over N_{\g}^{\rm rec}} \simeq 
0.96\times 10^{-2}\,N_{B-L}^{\rm lep,f} \,  ,
\ee
accounting for sphaleron conversion \cite{sphalerons} and photon dilution
and where, in the last numerical expression, we normalised the abundances $N_X$ of any generic quantity $X$  
in a way that the ultra-relativistic equilibrium abundance of a RH neutrino 
$N_{N_i}^{\rm eq}(T \gg M_i) =1$. 
Successful leptogenesis requires that $\eta_B^{\rm lep}$ 
reproduces the experimental value that, from {\em Planck} data (including lensing) combined with
external data sets \cite{planck15}, is given by
\be\label{etaBPlanck}
\eta_{B}^{CMB} = (6.10 \pm 0.04)\, \times 10^{-10}  \,   . 
\ee
For both two terms in Eq.~(\ref{2terms}) we can give analytic expressions. 
The relic value of the pre-existing asymmetry has to be calculated  \cite{problem,strongSO10,lbstrong} as $N_{B-L}^{\rm p, f} = \sum_{\a} \,  N_{\D_\a}^{\rm p,f}$,  with each flavour contribution given by
\bea\label{finalpas}
N_{\D_\t}^{\rm p,f} & = & 
(p^0_{{\rm p}\t}+\D p_{{\rm p}\tau})\,  e^{-{3\pi\over 8}\,(K_{1\t}+K_{2\t})} \, N_{B-L}^{\rm p,i} \,  , \\  \nonumber
N_{\D_\m}^{\rm p,f} & = & 
\left\{(1-p^0_{{\rm p}\t})\,\left[
p^0_{\mu\t_2^{\bot}}\, p^0_{{\rm p}\t^\bot_2}\,
e^{-{3\pi\over 8}\,(K_{2e}+K_{2\m})} + (1-p^0_{\m\t_2^{\bot}})\,(1-p^0_{{\rm p}\t^\bot_2}) \right]  + 
\D p_{{\rm p}\mu}\right\}
\,e^{-{3\pi\over 8}\,K_{1\m}}\, N_{B-L}^{\rm p,i}
 ,  \\  \nonumber
N_{\D_e}^{\rm p,f}& = & 
\left\{(1-p^0_{{\rm p}\t})\,\left[ 
p^0_{e\t_2^{\bot}}\,p^0_{{\rm p}\t^\bot_2}\,
e^{-{3\pi\over 8}\,(K_{2e}+K_{2\m})} + (1-p^0_{e \t_2^{\bot}})\,(1-p^0_{{\rm p}\t^\bot_2}) \right]  + \D p_{{\rm p} e}\right\}
 \,e^{-{3\pi\over 8}\,K_{1e}} \,\, N_{B-L}^{\rm p,i} \,   .
\eea
In this expression the $K_{i\a}$'s are the {\em flavoured decay parameters} defined by
\be
K_{i\a} \equiv {\G_{i\a}+\overline{\G}_{i\a}\over H(T=M_i)}= 
{|m_{D\a i}|^2 \over M_i \, m_{\star}} \,  ,
\ee
where  $\Gamma_{i\a}=\Gamma (N_i \ra \phi^\dagger \, l_\alpha)$ 
and $\bar{\Gamma}_{i \a}=\Gamma (N_i \ra \phi \, \bar{l}_\alpha)$ are the
zero temperature limit of the flavoured decay rates into $\a$ leptons
and anti-leptons in the three-flavoured regime, $m_{\star}\simeq 1.1 \times 10^{-3}\,{\rm eV}$
is the equilibrium neutrino mass, $H(T)=\sqrt{g^{SM}_{\star}\,8\,\pi^3/90}\,T^2/M_{\rm P}$ is the expansion rate
and $g_{\star}^{SM}=106.75$ is the SM number of ultra-relativistic degrees of freedom.
Using the bi-unitary parameterisation Eq.~(\ref{svd}), the flavoured decay parameters can be written as
\be\label{KialVL}
K_{i\a} = {\sum_{k,l} \, 
m_{Dk}\, m_{Dl} \,V_{L k\a} \, V_{L l \a}^{\star} \, U^{\star}_{R ki} \, U_{R l i} 
\over M_i \, m_{\star}}\,  .
\ee
The quantities $p^0_{{\rm p} \tau}$ and $p^0_{{\rm p}\tau_2^{\bot}}$ 
indicate the fractions of the pre-existing asymmetry in the tauon flavour and 
in the flavour $\tau_2^{\bot}$,  the  electron and muon flavours  superposition component 
in the leptons  produced by the $N_2$-decays (or equivalently the flavour component that is 
washed-out in the inverse processes producing $N_2$) so that $p^0_{{\rm p} \tau}+
p^0_{{\rm p}\tau_2^{\bot}} =1$.
The two quantities $p^0_{\alpha \t_2^{\bot}} \equiv K_{2\alpha}/(K_{2e}+K_{2\mu}) \; (\a=e,\mu)$ are
then the fractions of $\a$-asymmetry in the  $\tau_2^{\bot}$ component, so that
$p^0_{e \t_2^{\bot}}+p^0_{\mu \t_2^{\bot}} =1$.

The contribution from leptogenesis also has to be calculated as the sum of
three contributions from each flavour, explicitly 
\be
N_{B-L}^{\rm lep,f} = N_{\D_e}^{\rm lep,f} + N_{\D_\mu}^{\rm lep,f} + N_{\D_\tau}^{\rm lep,f} \,  .
\ee
The expression we derived for $M_1$ from the $SO(10)$ inspired conditions,
 the first of the Eqs.~(\ref{Mi}), implies $M_1 \ll 10^{9}\,{\rm GeV}$
and in this case the asymmetry produced from $N_1$ decays is negligible \cite{di,predictions}.
On the other hand $M_2$ can be sufficiently large 
\footnote{A lower bound $M_2 \gtrsim 5 \times 10^{10}\,{\rm GeV}$ was
found in \cite{riotto2}.} for the asymmetry produced from
$N_2$-decays to reproduce the observed asymmetry: for this
reason $SO(10)$-inspired conditions necessarily require a $N_2$-dominated scenario 
of leptogenesis.  
\footnote{In principle one should also consider the asymmetry produced 
from $N_3$ decays occurring in the unflavoured regime since $M_3 \gg 10^{12}\,{\rm GeV}$.
 However the $C\!P$ asymmetry $\ve_3$ is  suppressed as $M_2/M_3$ compared to 
 $\ve_2$ and in the end it turns out that also the contribution to the asymmetry 
 from $N_3$ decays, as that one from $N_1$ decays, is negligible.}
Moreover since just marginal solutions are found for $M_2 \gtrsim 10^{12}\,{\rm GeV}$,
where the production occurs in the unflavoured regime, one has to consider
a two-flavour regime for the asymmetry production from $N_2$ decays.
In this case the three flavoured asymmetries can be calculated using  \cite{vives,bounds,fuller,density}
\footnote{These equations for the calculation of the final asymmetry hold for 
$100 \,{\rm GeV} \lesssim  M_1 \lesssim 10^9\,{\rm GeV}$, in the $N_2$-dominated scenario. 
While the upper bound is basically
always valid within given $SO(10)$-inspired conditions, except for a very fine-tuned case 
corresponding to very small value of $\widetilde{m}_{\nu 11}$  (we will be back on this case), the lower bound in
principle could be violated if $\a_1 \lesssim 0.1$. In this case there is no wash-out from the
lightest RH neutrino and the exponentials would disappear and consequently all constraints on low energy neutrino
parameters. This scenario has been discussed in \cite{susy}.}
\bea\label{twofl} \nonumber
N_{\D_e}^{\rm lep,f} & \simeq &
\left[{K_{2e}\over K_{2\tau_2^{\bot}}}\,\ve_{2 \tau_2^{\bot}}\kappa(K_{2 \tau_2^{\bot}}) 
+ \left(\ve_{2e} - {K_{2e}\over K_{2\tau_2^{\bot}}}\, \ve_{2 \tau_2^{\bot}} \right)\,\kappa(K_{2 \tau_2^{\bot}}/2)\right]\,
\, e^{-{3\pi\over 8}\,K_{1 e}}  \,   , \\ \nonumber
N_{\D_\m}^{\rm lep,f} & \simeq & \left[{K_{2\mu}\over K_{2 \tau_2^{\bot}}}\,
\ve_{2 \tau_2^{\bot}}\,\kappa(K_{2 \tau_2^{\bot}}) +
\left(\ve_{2\mu} - {K_{2\mu}\over K_{2\tau_2^{\bot}}}\, \ve_{2 \tau_2^{\bot}} \right)\,
\kappa(K_{2 \tau_2^{\bot}}/2) \right]
\, e^{-{3\pi\over 8}\,K_{1 \mu}} \,  , \\
N_{\D_\t}^{\rm lep,f} & \simeq & \ve_{2 \tau}\,\kappa(K_{2 \tau})\,e^{-{3\pi\over 8}\,K_{1 \tau}} \,  ,
\eea
where  $\ve_{2\a} \equiv -(\G_{2\a}-\overline{\G}_{2\a})/(\G_2 + \overline{\G}_2)$ are the $N_2$-flavoured $C\!P$ asymmetries $(\a =e, \mu, \t)$, with $\G_2 \equiv \sum_{\a} \G_{2\a}$ and
$\overline{\G}_{2}\equiv \sum_{\a} \, \overline{\G}_{2\a}$, and 
simply $\ve_{2 \tau_2^{\bot}} \equiv \ve_{2e}+\ve_{2\m}$ and 
$K_{2 \tau_2^{\bot}} \equiv K_{2e}+K_{2\m}$.
For the efficiency factors at the production $\k(K_{2\a})$
we used the standard analytic expression \cite{predictions}
\be\label{kappa}
\k(K_{2\a}) = 
{2\over z_B(K_{2\a})\,K_{2\a}}
\left(1-e^{-{K_{2\a}\,z_B(K_{2\a})\over 2}}\right) \,  , \;\; z_B(K_{2\a}) \simeq 
2+4\,K_{2\a}^{0.13}\,e^{-{2.5\over K_{2\a}}} \,   .
\ee
This expression holds for an initial thermal abundance but since 
all solutions we  found are for strong wash-out at the production
(either $K_{2\t} \gg 1$ or $K_{2 \tau_2^{\bot}} \gg 1$ respectively for tauon and muon-dominated solutions),
the asymmetry does not depend on the initial $N_2$ abundance anyway.
Moreover in the strong wash-out regime the theoretical uncertainties are within $20\%$.\cite{upperbound,bodeker}
\footnote{Notice that since the constraints on the low energy neutrino parameters 
are determined mainly by the vanishing of the $K_{1\a}$ in the exponentials,
these depend only logarithmically on the asymmetry and a theoretical uncertainty
of  $20\%$ on the asymmetry translates into a less than $1\%$ theoretical uncertainty on the constraints.
In any case improvements in this direction will also be needed in future.}

The flavoured $C\!P$ asymmetries can be calculated using \cite{crv}
\be\label{eps2a}
\ve_{2\a} \simeq
\overline{\ve}(M_2) \, \left\{ {\cal I}_{23}^{\a}\,\x(M^2_3/M^2_2)+
\,{\cal J}_{23}^{\a} \, \frac{2}{3(1-M^2_2/M^2_3)}\right\}\, ,
\ee
where we introduced
\be
\overline{\ve}(M_2) \equiv {3\over 16\,\pi}\,{M_2\,m_{\rm atm} \over v^2} \, , \hspace{3mm} \xi(x)=\frac{2}{3}x\left[(1+x)\ln\left(\frac{1+x}{x}\right)-\frac{2-x}{1-x}\right] \,  ,
\ee
\be
{\cal I}_{23}^{\a} \equiv   {{\rm Im}\left[m_{D\a 2}^{\star}
m_{D\a 3}(m_D^{\dag}\, m_D)_{2 3}\right]\over M_2\,M_3\,\mtt\,m_{\rm atm} }\,   
\hspace{5mm}
\mbox{\rm and}
\hspace{5mm}
{\cal J}_{23}^{\a} \equiv  
{{\rm Im}\left[m_{D\a 2}^{\star}\, m_{D\a 3}(m_D^{\dag}\, m_D)_{3 2}\right] 
\over M_2\,M_3\,\mtt\,m_{\rm atm} } \,{M_2\over M_3}   ,
\ee
with $\mtt \equiv (m_D^{\dag}\, m_D)_{2 2}/M_2$.
Since $M_3 \gg M_2$,  
one can approximate $\xi(M_3^2/M_2^2) \simeq 1$
and neglect the second term $\propto {\cal J}^{\a}_{23}$ in the Eq.~(\ref{eps2a}). 
\footnote{This hierarchical approximation has been tested since we used
the exact expression for $\xi$, able to describe a resonant enhancement,  
and we did not neglect the term ${\cal J}_{23}^{\a}$ in the numerical results and
we checked that no new quasi-degenerate solutions are found 
for $M_1 \lesssim 10^9\,{\rm GeV}$ (i.e. within the $N_2$-dominated scenario) \cite{strongSO10}. 
We will come back on this point 
when we will discuss theoretical uncertainties at the end of this Section.}
Using the singular value decomposition Eq.~(\ref{svd}) for $m_D$, one obtains 
\be\label{ve2alAN}
\ve_{2\a} \simeq {3 \over 16\, \pi \, v^2}\,
{|(\widetilde{m}_{\nu})_{11}| \over m_1 \, m_2 \, m_3}\,
{\sum_{k,l} \, m_{D k} \, m_{Dl}  \, {\rm Im}[V_{L k \a }  \,  V^{\star}_{L  l \a } \, 
U^{\star}_{R k 2}\, U_{R l 3} \,U^{\star}_{R 3 2}\,U_{R 3 3}] 
\over |(\widetilde{m}_{\nu}^{-1})_{33}|^{2} + |(\widetilde{m}_{\nu}^{-1})_{23}|^{2}}   \,  ,
\ee
where  in $(m^{\dagger}_D\,m_D)_{22} = \sum_k \, m^2_{Dk} \, |U_{R k2}|^2$ we neglected
the term $k=1$, suppressed as  $(m_{D1}/m_{D2})^2$ compared to the others,
and we have also approximated 
$(m^{\dagger}_D\,m_D)_{23} \simeq  m^2_{D3} \, U^{\star}_{R 3 2}\,U_{R 3 3}$.

Except for special points where $\ve_{2e} \simeq \ve_{2\m}$, one of the two (typically $\ve_{2\m}$)
dominates on the other and this implies that the  terms in 
$N_{\D_e}^{\rm lep,f}$ and $N_{\D_\mu}^{\rm lep,f}$ in round brackets, the so called phantom terms \cite{fuller,density}, 
are necessarily negligible and one obtains much simpler expressions, 
\footnote{Like for the hierarchical approximation, we indeed also checked that phantom terms, that are kept
in numerical results, do not play any role and can be neglected in $SO(10)$-inspired leptogenesis.}
\bea\label{twofl} \nonumber
N_{\D_e}^{\rm lep,f} & \simeq &
\ve_{2e} \, \kappa(K_{2e} + K_{2\m}) \,
\, e^{-{3\pi\over 8}\,K_{1 e}}  \,   , \\ \nonumber
N_{\D_\m}^{\rm lep,f} & \simeq & 
\ve_{2\m}\,\kappa(K_{2e} + K_{2\m}) 
\, e^{-{3\pi\over 8}\,K_{1 \mu}} \,  , \\
N_{\D_\t}^{\rm lep,f} & \simeq & \ve_{2 \tau}\,\kappa(K_{2 \tau})\,e^{-{3\pi\over 8}\,K_{1 \tau}} \,  ,
\eea
where we wrote again  $N_{\D_\t}^{\rm lep,f}$ for completeness. In this way, using the analytic
expression Eq.~(\ref{UR}) for $U_R$,  Eq.~(\ref{ve2alAN}) for the $\ve_{2\a}$'s and Eq.~(\ref{KialVL}) for the $K_{i\a}$'s, and given an expression for $V_L$, one obtains a full analytical expression for the asymmetry depending on the low energy neutrino parameters, on $V_L$ and on the $\a_i$'s 
\footnote{Notice however that the dependence
on $\a_1$ and on $\a_3$ cancels out in physical solutions satisfying successful leptogenesis \cite{riotto2,SO10decription}).}. 
In the Appendix we summarise all this set of analytic expressions that basically constitute the
analytical solution we found.

If one adopts the approximation $V_L = I$, 
from the  Eq.~(\ref{ve2alAN})  one obtains 
for the $\ve_{2\a}$'s \cite{SO10decription}
\be\label{ve2alVLI}
\ve_{2\a} \simeq {3\,m^2_{D\a} \over 16\, \pi \, v^2}\,
{m_{ee} \over m_1 \, m_2 \, m_3}\,
{{\rm Im}[U^{\star}_{R \a 2}\,U_{R \a 3}\,U^{\star}_{R 3 2}\,U_{R 3 3}] 
\over |(m_{\nu}^{-1})_{\t\t}|^{2} + |(m_{\nu}^{-1})_{\m\t}|^{2}}   \,  .
\ee
From this one it is then easy to obtain explicitly
\bea\label{ve2tauVLI}
\ve_{2\t} & \simeq & {3\, m^2_{D 2} \over 16\,\pi\,v^2}\,  {m_{ee}\,
\left[|(m_{\nu}^{-1})_{\t\t}|^{2} + |(m_{\nu}^{-1})_{\m\t}|^{2} \right]^{-1} \over m_1\,m_2\,m_3 \, 
}\,
{|(m^{-1}_{\n})_{\m\t}|^2 \over |(m_{\nu}^{-1})_{\t\t}|^{2}} 
\,\sin\a_L    \, ,    \\  \nonumber
\ve_{2\mu} 
& \simeq & - {m^2_{D2} \over m^2_{D3}} \, \ve_{2\t} \, ,  \\ \nonumber
\ve_{2e} & \simeq &  {3\,m^2_{D1}\over 16\,\pi \, v^2}\, {m^2_{D1} \over m^2_{D3}}\, {
|m_{\n e\m}|\, \left[|(m_{\nu}^{-1})_{\t\t}|^{2} + |(m_{\nu}^{-1})_{\m\t}|^{2} \right]^{-2}
\over m_1\,m_2\,m_3 }\,
{|(m^{-1}_{\n})_{e\t}|\, |(m^{-1}_{\n})_{\m\t}| \over |(m^{-1}_{\n})_{\t\t}|^{2}} \, 
\sin\a^e_L  \,  ,
\eea
implying
\be\label{ratiosVLI}
\ve^{\rm max}_{2\t}:\ve^{\rm max}_{2\m}:\ve^{\rm max}_{2e} \sim  1 : {m^2_{D2} \over m^2_{D3}} :  
{m^2_{D2} \over m^2_{D3}} \, {m^4_{D1} \over m^4_{D2}}  \,   ,
\ee
where we maximised over the phase factors given by
\be
\a_L =  {\rm Arg}\left[m_{\nu ee}\right]  - 2\,{\rm Arg}[(m^{-1}_{\nu})_{\m\t}] - \pi -2\,(\rho+\s)  \,  ,
\ee
and
\be
\a^e_L =  {\rm Arg}\left[m_{\nu e \m}\right]  - {\rm Arg}[(m^{-1}_{\nu})_{\m\t}] 
- {\rm Arg}[(m^{-1}_{\nu})_{ e \t}] - \pi -2\,(\rho+\s)  \,   .
\ee
The electron $C\! P$ asymmetry is so strongly suppressed 
(more than fifteen orders of magnitude  compared to the tauonic)
that the corresponding contribution to the final asymmetry is completely negligible. 
The muon $C\!P$ asymmetry is also suppressed compared to the tauonic
$C\!P$ asymmetry by about four orders of magnitude but it might be still large enough to allow 
the existence of (marginal) muon-dominated solutions. 
However, when the wash-out both at the production and from the lightest RH neutrino
is also taken into account, one finds that also the muon contribution to the final asymmetry is always 
much below the observed value and one does not find any muon-dominated solution for $V_L =I$ \cite{riotto2}.
Therefore, the electron and muon contributions are never able to reproduce the observed asymmetry 
and the final asymmetry can be approximated just by the tauon contribution, 
so that we can write \cite{SO10decription,susy}
\bea\label{NBmLfVLI}
\left. N_{B-L}^{\rm lep,f} \right|_{V_L =I} & \simeq & 
{3\over 16\,\pi}\, {m_{D2}^2 \over v^2}\, {|m_{\nu ee}|\,
(|m^{-1}_{\nu \t \t}|^2 + |m^{-1}_{\nu \m \t}|^2)^{-1} \over m_1\,m_2\,m_3}\,
{|m^{-1}_{\n\m\t}|^2\over |m^{-1}_{\n\t\t}|^2}\,\sin\a_L    \\  \nonumber
& \times & \kappa\left({m_1\,m_2\,m_3 \over m_{\star}}\, 
{|(m_{\nu}^{-1})_{\m \t}|^2 \over |m_{\nu ee}|\, |(m_{\nu}^{-1})_{\t \t}|} \right)  \\  \nonumber
& \times & 
e^{-{3\pi\over 8}\,{|m_{\nu e\t}|^2 \over m_{\star}\,|m_{\nu ee}|}  }  
\,  ,
\eea
where for the $K_{i\a}$'s we used \cite{SO10decription}
\be\label{KialVLeqI}
K_{i\a} =   {m^2_{D\a}  \over M_i \, m_{\star}} \, |U_{R \a i}|^2  \,  ,
\ee
that can be easily derived from the Eq.~(\ref{KialVL}) for $V_L = I$,  obtaining
\be\label{K1tauVLI}
K_{1\t}   \simeq  {m^2_{D3} \over m_{\star} \, M_1} \, |U_{R 3 1}|^2  
\simeq {|m_{\nu e\t}|^2 \over m_{\star}\,|m_{\nu ee}|}  
\ee
and 
\be\label{K2tauVLI}
K_{2\t} \simeq {m^2_{D3} \over m_{\star} \, M_2} \, |U_{R 3 2}|^2  
\simeq {m_1\,m_2\,m_3 \over m_{\star}}\, 
{|(m_{\nu}^{-1})_{\m \t}|^2 \over |m_{\nu ee}|\, |(m_{\nu}^{-1})_{\t \t}|}  \,   .
\ee

In Fig.~1 we show scatter plots of solutions for successful $SO(10)$-inspired leptogenesis in the seesaw
parameter space projected on planes for different choices of two low energy neutrino parameters. We distinguish in yellow and orange the tauon-dominated solutions, corresponding respectively to  $I \leq V_L \leq V_{CKM}$  and  $V_L =I$,  and in green the muon-dominated solutions realised only for $I \leq V_L \leq V_{CKM}$.
The variation of the low energy neutrino parameters is within the indicated  
ranges.
\footnote{To be conservative we used $4\s$ intervals in Eq.~(\ref{expranges}) and 
$\d$  is allowed to vary in the whole range $[-\pi,\pi]$.} 
The plots have been  obtained for  $(\a_1,\a_2,\a_3)=(1,5,1)$. We have also used 
\footnote{Since the final asymmetry $ \propto m_{D2}^2=(\a_2\,m_{\rm c})^2$, 
for a given value of $\a_2$, the theoretical uncertainty in the 
 determination of the value of $m_{\rm c}$ at the leptogenesis scale
 translates into a (doubled) theoretical uncertainty in the determination of the final asymmetry: 
 the value of the charm quark  mass at the scale of leptogenesis is then one of the most important
 sources of uncertainties.}
$(m_{\rm u},m_{\rm c}, m_{\rm t})=(1\,{\rm MeV}, 400\,{\rm MeV}, 100\,{\rm GeV})$
for the values of the up quark masses at the leptogenesis scale $T_{L} \simeq (3$--$10)\,10^{10}\,{\rm GeV}$ \cite{quarks}.

We also show the subset of solutions satisfying in addition the strong thermal condition 
(light blu for $I \leq V_L \leq V_{CKM}$ and dark blue for $V_L =I$) for an initial
pre-existing asymmetry $N^{\rm p,i}_{B-L}=10^{-3}$.
The scatter plots have been obtained for an initial thermal $N_2$ abundance
but, as we will discuss, the solutions do not depend on the initial value of $N_{N_2}$.
We have also imposed $M_{\O} \equiv {\rm max}_{i,j}[|\O_{ij}^2|] =100$. 
With these conditions we haven't found any electronic-dominated solution,
even for $I \leq V_L \leq V_{CKM}$, we will be back on this point. 
\begin{figure}
\begin{center}
\psfig{file=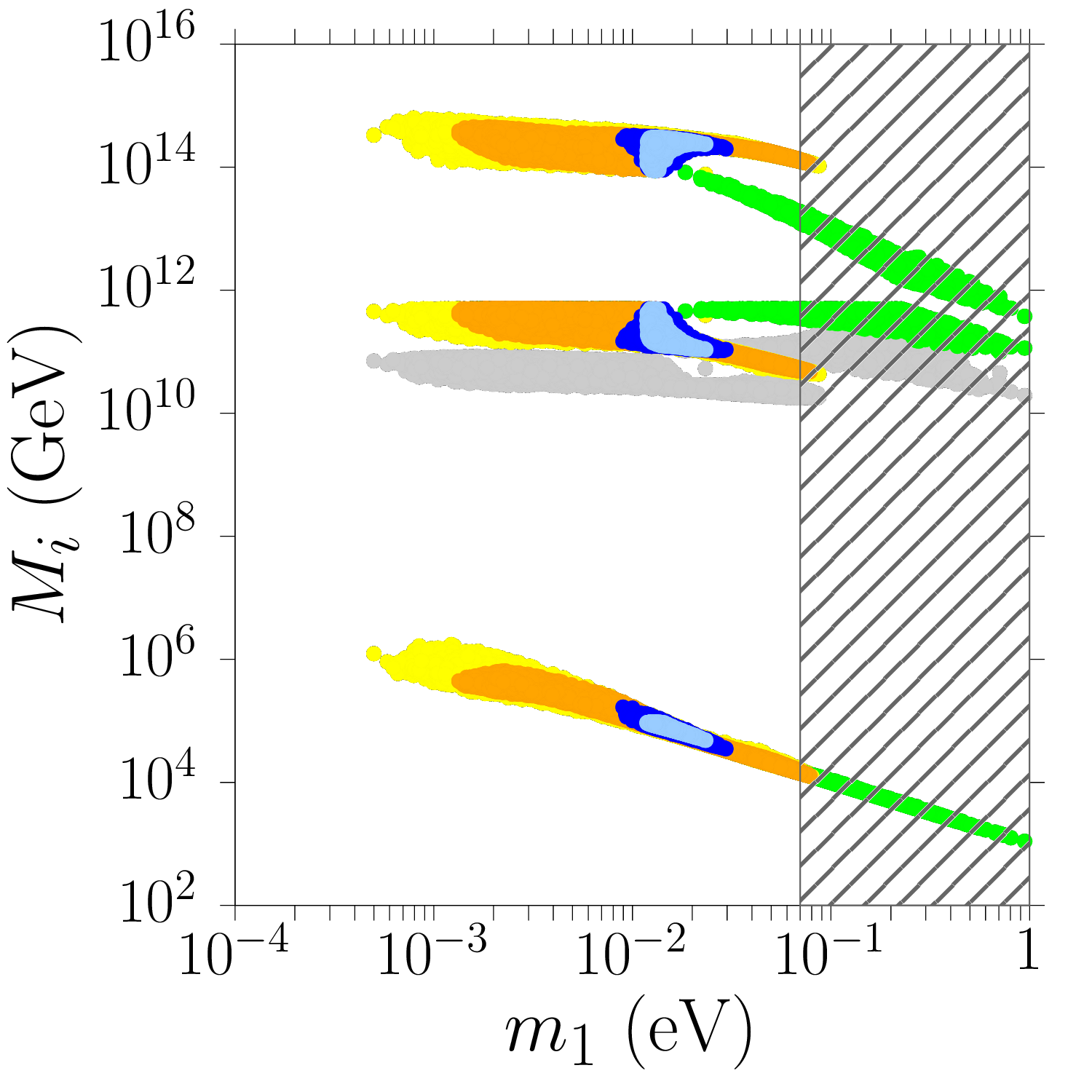,height=50mm,width=52mm}
\hspace{-1mm}
\psfig{file=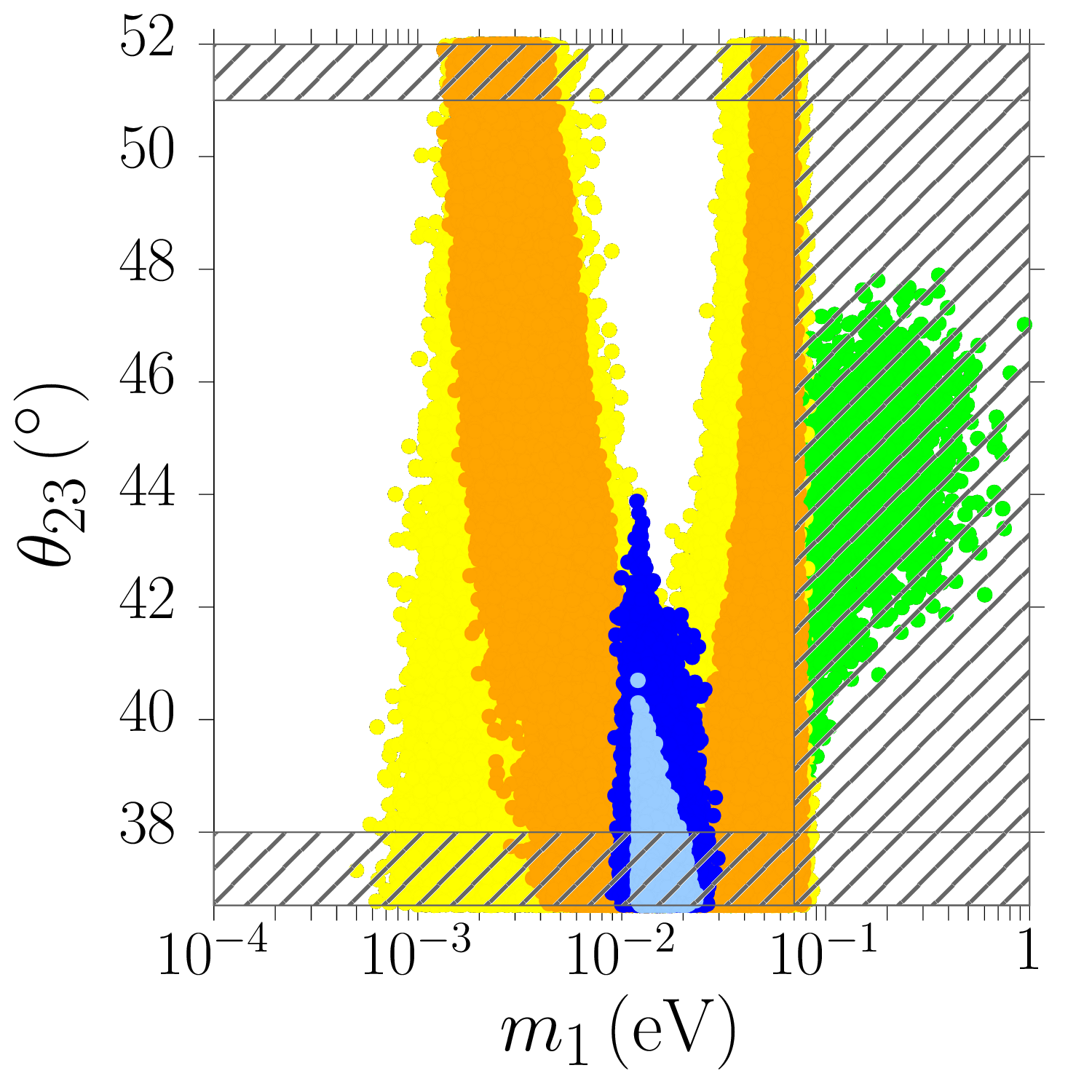,height=50mm,width=52mm}
\hspace{-1mm}
\psfig{file=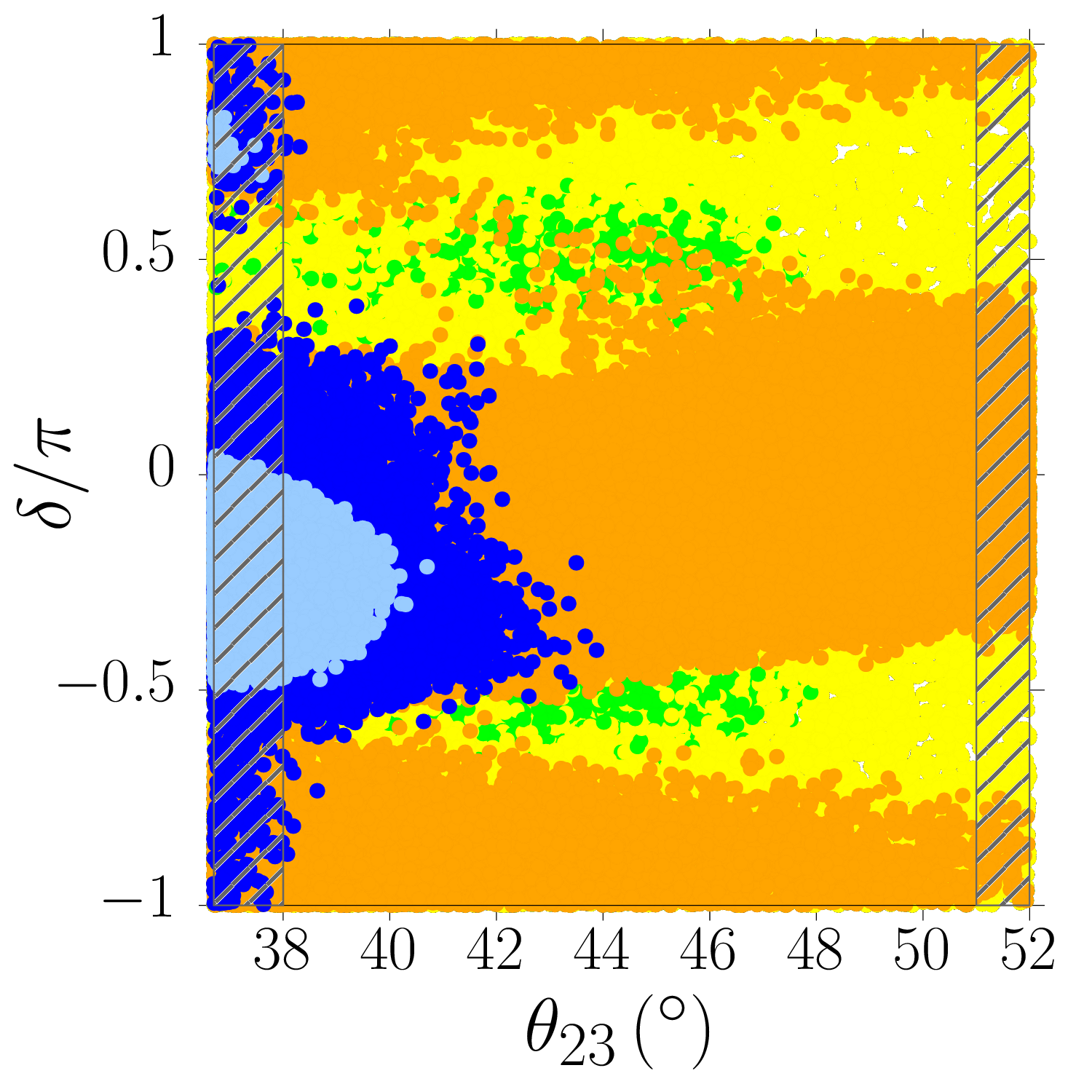,height=50mm,width=52mm} \\
\psfig{file=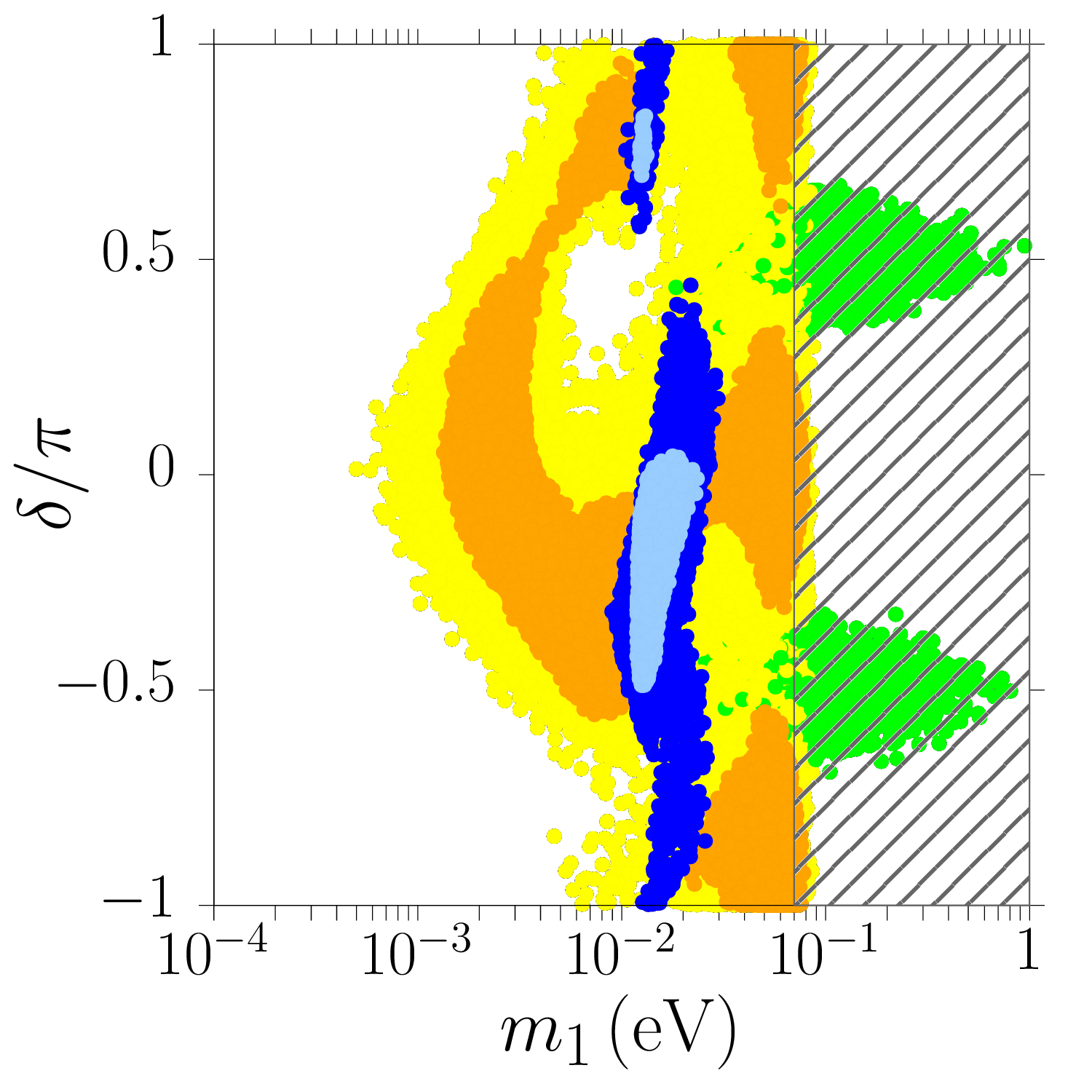,height=50mm,width=52mm}
\hspace{-1mm}
\psfig{file=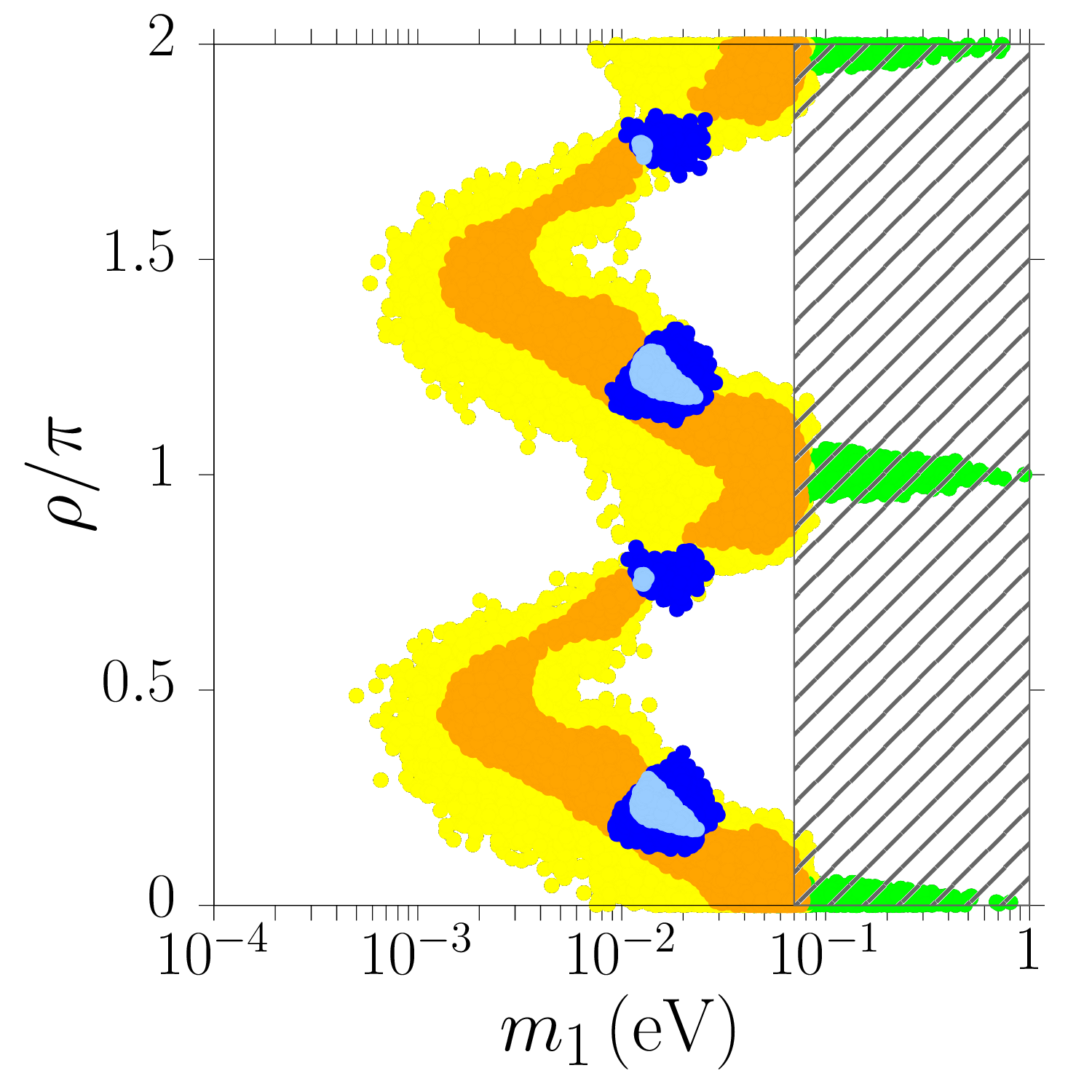,height=50mm,width=52mm}
\hspace{-1mm}
\psfig{file=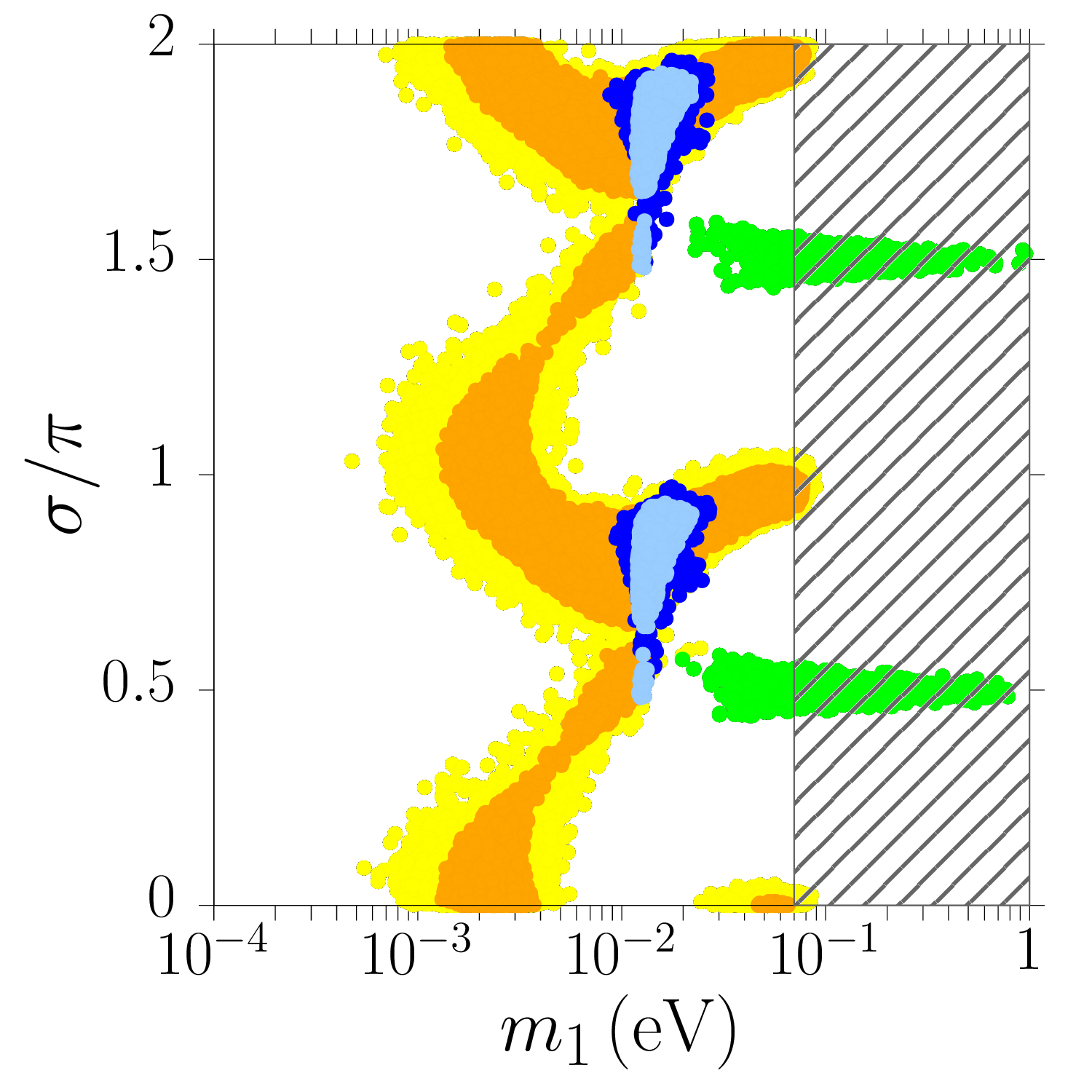,height=50mm,width=52mm} \\
\psfig{file=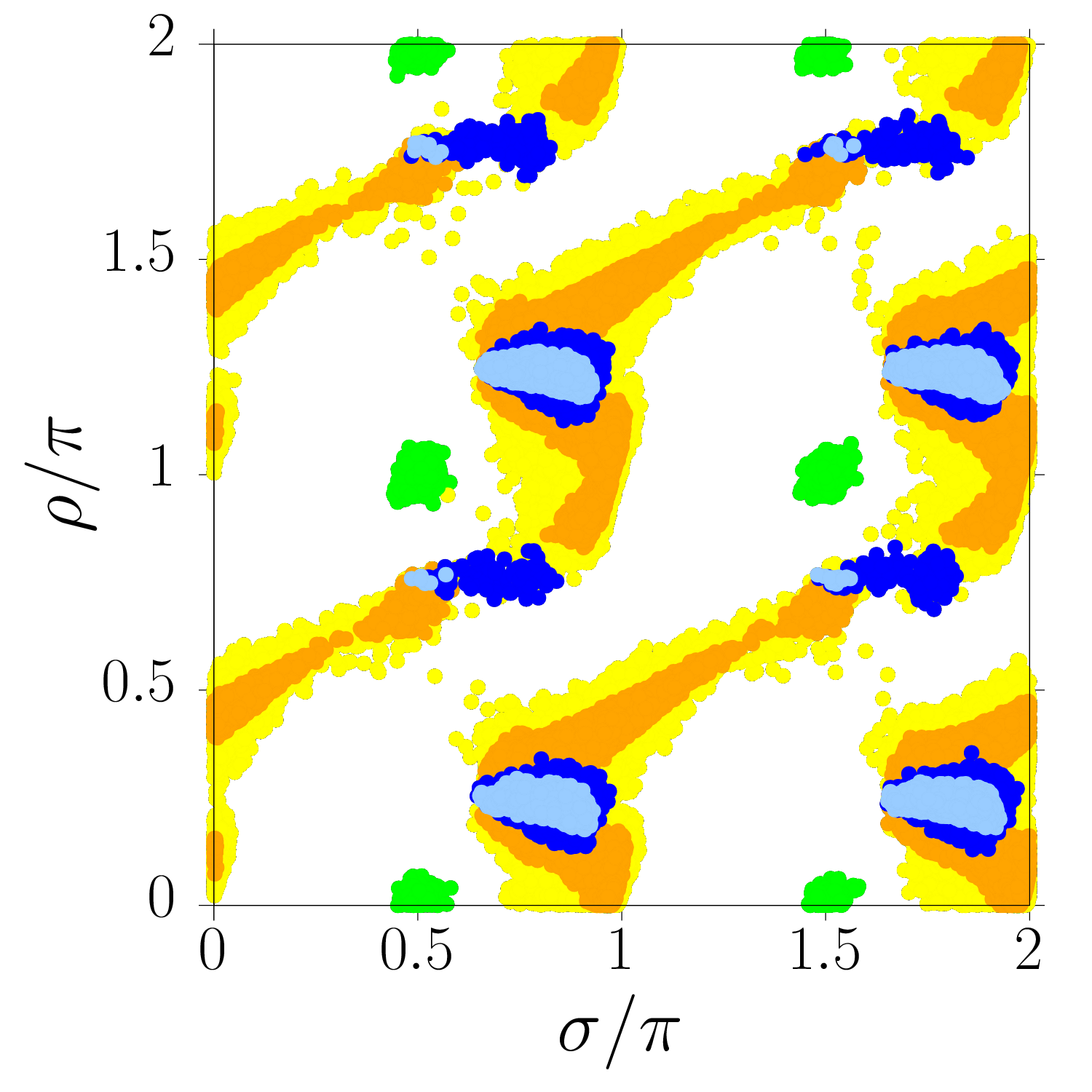,height=50mm,width=52mm}
\hspace{-1mm}
\psfig{file=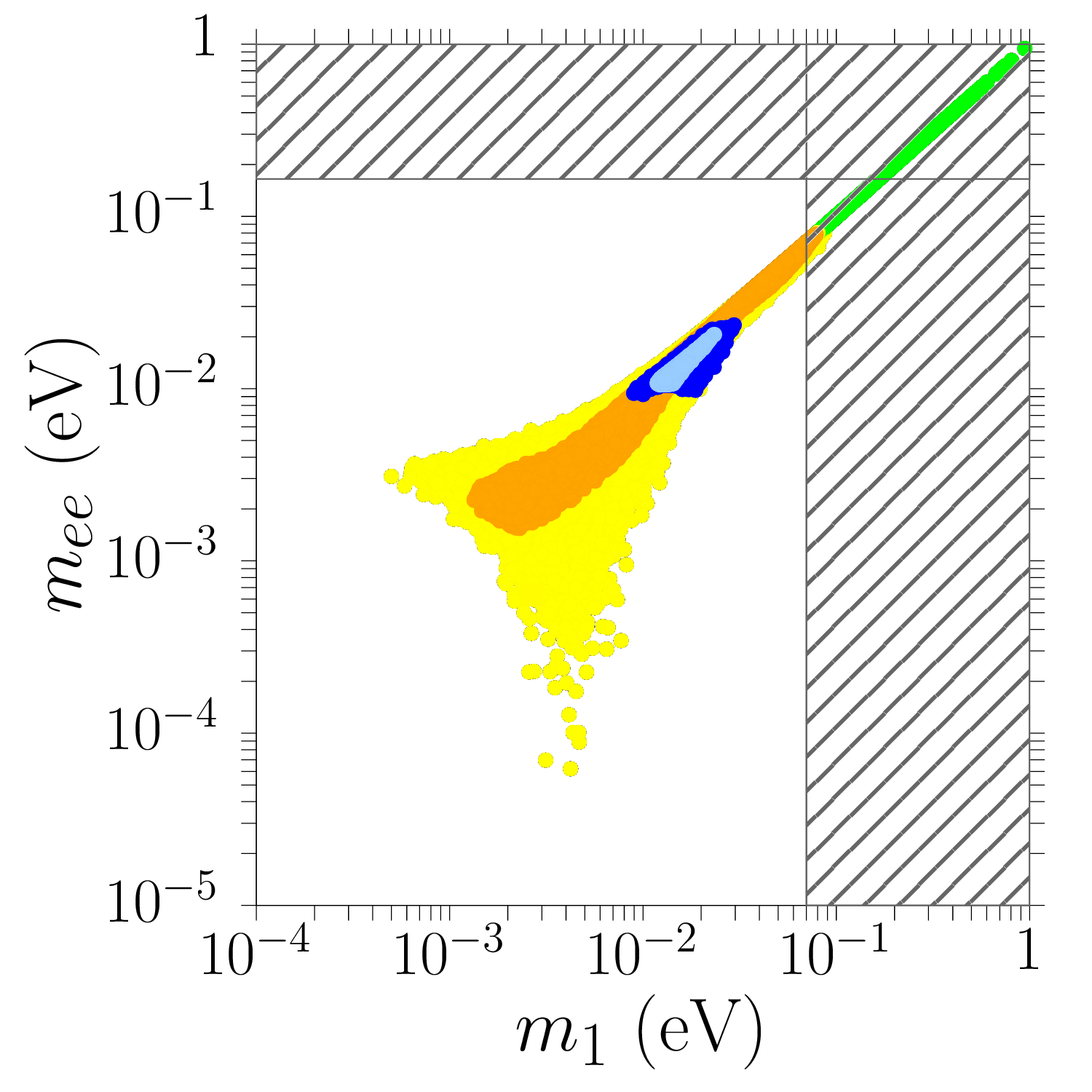,height=50mm,width=52mm}
\hspace{-1mm}
\psfig{file=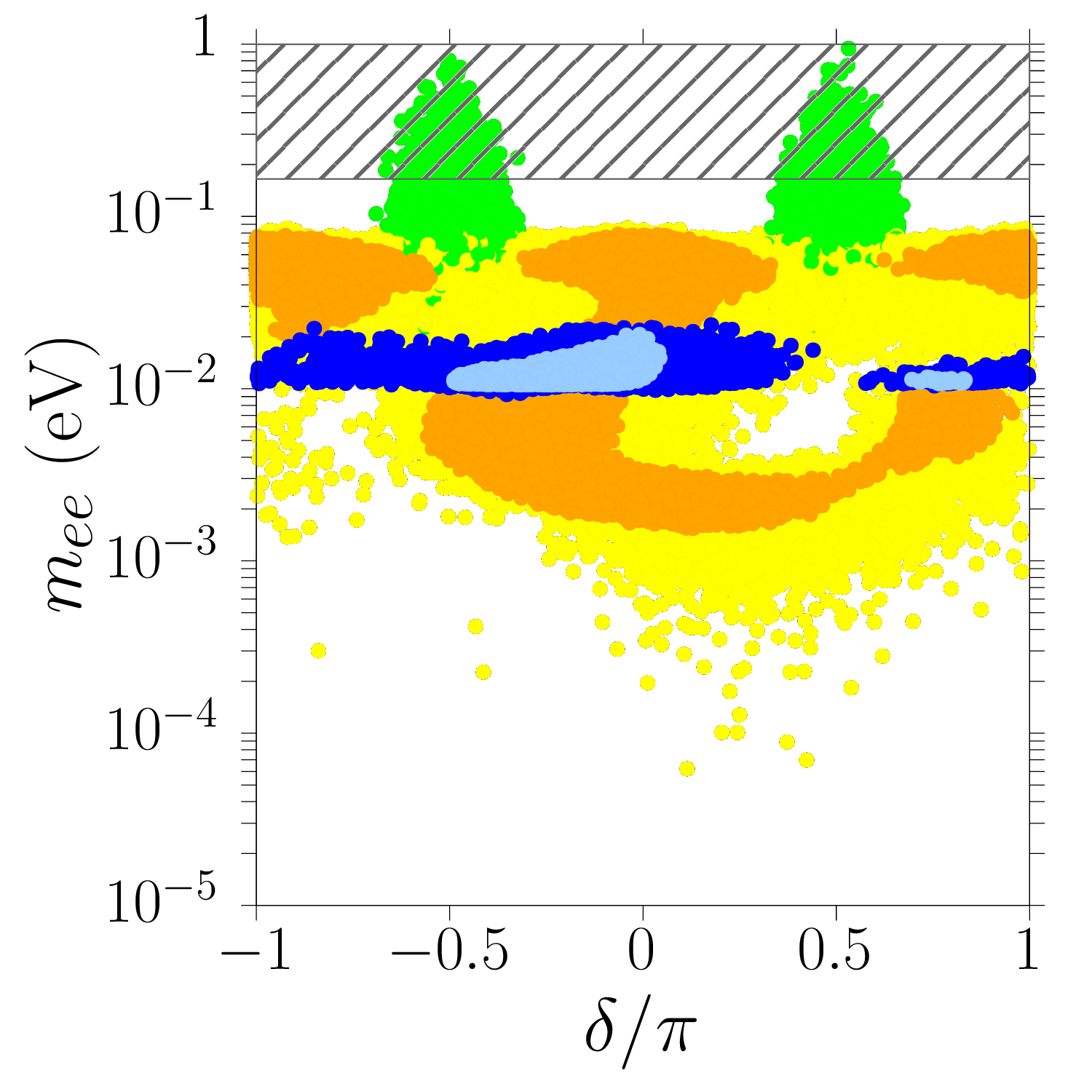,height=50mm,width=52mm} 
\end{center}
\vspace{-5mm}
\caption{Scatter plots in the seesaw parameter space projected on 
different planes for NO and  $(\a_1,\a_2,\a_3 =1,5,1)$. 
All points satisfy  (at $\simeq 3\s$) successful leptogenesis. 
 The yellow points correspond to tauon-dominated solutions for an initial
vanishing pre-existing asymmetry (light yellow 
for $I \leq V_L \leq V_{CKM}$ and orange for $V_L =I$)
 The blue points are the subset satisfying the additional strong thermal condition for an initial value 
of the  pre-existing asymmetry $N_{B-L}^{\rm p,i}=10^{-3}$
(dark blue for $I \leq V_L \leq V_{CKM}$ and light blue for $V_L =I$).
The green points correspond to muon-dominated solutions. 
The solutions have been obtained for $M_{\Omega}=100$ and imposing $M_3 > 2\,M_2$.
The dashed bands indicate the $3\s$ excluded ranges for the corresponding mixing parameters
(see Eq.~(\ref{expranges})), while for $m_1$ and $m_{ee}$ they indicate the $95\%$C.L. and $90\%$C.L.
upper bounds from {\em Planck} and KamLAND-Zen.}
\label{constrNO}
\end{figure}
Contrarily to $\a_1$ and $\a_3$ that cancel out in the final asymmetry, 
the parameter $\a_2 \equiv m_{D2}/m_{\rm c}$ plays clearly a very important role
since all $N_2$ $C\!P$ asymmetries are proportional to the square of this parameter. 
We have set $\a_2 =5$ as maximum reference value. The dependence of the constraints
on $\a_2$ was studied in detail in \cite{riotto2} where the lower bound 
$\a_2 \gtrsim 1$ was found. There is no of course upper bound  from leptogenesis but in realistic 
models this is never found too much larger than our reference value $\a_2 =5$. For example it is interesting that
in the realistic fits found in \cite{rodejohann} within $SO(10)$ models one has $\a_2 \lesssim 6$, very close to our reference maximum value. 

Before concluding this Section we want to comment on the approximations of our results that might
give rise to some corrections that should be therefore considered sources of theoretical uncertainties 
in our calculation. 

\begin{itemize}
\item We are using Boltzmann equations and for this reason we have imposed 
$M_2 \lesssim 10^{12} \,{\rm GeV}$, where a two-flavour regime is realised
at the $N_2$ production and Boltzmann equations can be used. If $M_2 \gg 10^{12}\,{\rm GeV}$
the production would occur in the unflavoured regime and the wash-out at production would be 
much higher and indeed if one also calculates the asymmetry in this regime
one finds very marginal points as discussed in detail in \cite{strongSO10}. 
One could therefore wonder whether in a density matrix approach, describing
the transition between the two regimes, one could find a suppression of solutions
already between $10^{11}\,{\rm GeV}$ and $10^{12}\,{\rm GeV}$.
However, we should say  that for $\a_2 < 5$ 
and $M_{\O} < 100$, as we are setting, in any case one does not find $M_2$ to be much larger than 
$10^{11}\,{\rm GeV}$ for most points. 
Therefore, we do not expect much more stringent constraints from
a density matrix formalism for $\a_2 \lesssim 5$. 
\item We are using a hierarchical approximation $M_3 \gtrsim M_2$.  However checks in \cite{strongSO10}
have found new quasi-degenerate solutions only at very large $m_1$ values, $m_1 \gtrsim 1\,{\rm eV}$, anyway
cosmologically excluded.  We are  neglecting just the case
of a compact spectrum $M_1 \sim M_2 \sim M_3 \sim 10^{10}\,{\rm GeV}$ realised
when both $|\widetilde{m}_{\nu 11}|$ and $|(\widetilde{m}_{\nu}^{-1})_{33}|$ get sufficiently small. 
\footnote{The reason why solutions in the vicinity of the crossing level $M_2 \sim M_3 \sim 10^{13}\,{\rm GeV}$,
realised when only $|(\widetilde{m}_{\nu}^{-1})_{33}|$ tends to vanish is that in this case 
$K_{2\tau} \propto |(\widetilde{m}_{\nu}^{-1})_{33}|^{-1}$ tends to become huge and together with it of course $K_2$ so that at the production one has a very strong wash-out and even the resonant enhancement of the
$C\!P$ asymmetries does not help. Of course in addition in any case one would also have huge fine-tuning
in the seesaw formula. The crossing level for which $M_1 \sim M_2 \sim 10^7 \,{\rm GeV}$
when $|\widetilde{m}_{\nu 11}|$ vanish is also excluded since in this case, even worse,  $K_{1\t} \propto |\widetilde{m}_{\nu 11}|^{-1}$. For this reason only a compact spectrum is left as a (very fine-tuned) caveat to a hierarchical spectrum.} The conditions
for this compact spectrum were studied in \cite{afs} and more recently, including flavour effects and within a
realistic model, in \cite{nardibuccella}.  However, as already noticed in \cite{SO10decription}, this special
case necessarily implies a huge fine-tuning in the seesaw formula. 
\footnote{It should be clear that the fine tuning is not only at the level of choosing the correct
value of the degeneracy to realise the right asymmetry, but also more seriously, 
as already noticed in the footnote 8, from a comparison of the expressions    Eq.~(\ref{Mi}) with  Eq.~(\ref{Omegaapp}) for $\O$, at the level of the seesaw formula.} 
In any case a compact spectrum solution gives rise to a distinct set of constraints on low energy neutrino parameters \cite{nardibuccella}, in particular it predicts no signal in $0\nu\b\b$ experiments since $m_{ee} \lesssim 1\,{\rm meV}$, while in our case the bulk of the solutions implies a detectable signal despite NO.

\item We are neglecting the running of the parameters (including a precise evaluation
of the charm quark mass at the scale of leptogenesis). This is not expected to be able to change significantly our results but of course in a not too far future, with an increase of the experimental precision on $\theta_{23}$ and $\d$ it might become necessary to include the running from radiative corrections. 
 
\item Another approximation we are using, and that might be a source of theoretical uncertainties, is that we are neglecting flavour coupling \cite{fuller}. This  generates new terms in the asymmetry (though usually sub-dominant) that can open new solutions and relax the constraints. An example was found in \cite{A2Zlep}, though the solution  was also requiring some large amount of fine tuning in the seesaw formula.  On the basis 
of preliminary results, we can say that flavour coupling introduces only corrections to the
analytical expression we found or it adds solutions involving great amount of fine-tuning in the see-saw formula \cite{preparation}. 
\end{itemize}

Our analytic solution will be actually very useful for a future derivation of the constraints 
including these effects, since it provides a new tool to generate solutions in a much faster way
and likely also to understand analytically the impact of the various effects.  

\section{Decrypting the impact of $V_L \simeq V_{CKM}$}

In this Section we want to understand the impact of turning on $V_L \simeq V_{CKM}$ using the 
analytical expressions obtained in the previous Section.
Since $V_L$ is unitary, this can be parameterised analogously to the leptonic 
mixing matrix as
\begin{equation}\label{Umatrix}
V_L=
\left( \begin{array}{ccc}
c^L_{12}\,c^L_{13} & s^L_{12}\,c^L_{13} & s^L_{13}\,e^{-{\rm i}\,\d_L} \\
-s^L_{12}\,c^L_{23}-c^L_{12}\,s^L_{23}\,s^L_{13}\,e^{{\rm i}\,\d_L} &
c^L_{12}\,c^L_{23}-s^L_{12}\,s^L_{23}\,s^L_{13}\,e^{{\rm i}\,\d_L} & s^L_{23}\,c^L_{13} \\
s^L_{12}\,s^L_{23}-c^L_{12}\,c^L_{23}\,s^L_{13}\,e^{{\rm i}\,\d_L}
& -c^L_{12}\,s^L_{23}-s^L_{12}\,c^L_{23}\,s^L_{13}\,e^{{\rm i}\,\d_L}  &
c^L_{23}\,c^L_{13}
\end{array}\right)
\, {\rm diag}\left(e^{i\,\rho_L}, 1, e^{i\,\sigma_L}  \right)\, ,
\end{equation}
having introduced three mixing angles $\theta_{12}^L, \theta_{13}^L$ 
and $\theta_{23}^L$ ($s^L_{ij} \equiv \sin \theta_{ij}^L$ and $c_{ij} \equiv \cos\theta_{ij}^L$),
one Dirac-like phase $\delta_L$ and two Majorana-like phases $\rho_L$ and $\sigma_L$.
We want to  understand analytically the effects of non-vanishing $\theta_{ij}^L$ with values
at the level of the respective angles in $V_{CKM}$ (see footnote 1), 
effects that have been  so far found only numerically.  

As we said, we will focus on NO, since for IO  the allowed regions,  in the non-supersymmetric framework we
are considering and for $\a_2 \lesssim 5$, are marginal 
(in particular they require necessarily $\theta_{23}$ in the second octant and $m_1 \gtrsim 10\,{\rm meV}$
implying $\sum_i m_i \gtrsim 0.11 \,{\rm eV}$, slightly disfavoured by current cosmological observations) and they completely disappear in the case of strong thermal leptogenesis. 
\footnote{In the supersymmetric case, for large $\tan\beta \gtrsim 15$, one can have solutions for
successful strong thermal leptogenesis even for IO \cite{susy}.}
These effects can be summarised as follows: 
\begin{itemize}
\item Turning on $V_L \simeq V_{CKM}$ enlarges the allowed region for tauon-dominated solutions
relaxing the constraint on the low energy neutrino parameters. 
There are two interesting features that should be understood with an analytic description: 
\begin{itemize}
\item[(i)] Within tauon-dominated solutions, there is a subset of solutions satisfying also the strong thermal
condition \cite{strongSO10,SO10decription}. As one can see from the blue regions in the top central panel in Fig.~1,  this subset is characterised by an upper bound on 
$\theta_{23}$ that for $V_L = I$ is given by $\theta_{23} \lesssim 41^{\circ}$ and for
$I \leq V_L \lesssim V_{CKM}$ relaxes to $\theta_{23} \lesssim 44^{\circ}$. The allowed range for $\d$
also enlarges for a given value of $\theta_{23}$. In the light
of the current best fit value for $\theta_{23} \simeq 41^{\circ}$ (see Eq.~(\ref{expranges})),
this is an interesting effect of turning on $V_L \simeq V_{CKM}$ to be understood.
\item[(ii)] the lower bound $m_{ee} \gtrsim 10^{-3}\,{\rm eV}$ strongly relaxes to 
$m_{ee} \gtrsim 5 \times 10^{-5}\,{\rm eV}$.
\end{itemize}
\item While for $V_L = I$ there are only tauon-dominated solutions able to reproduce the observed 
asymmetry \cite{riotto1}, muon-dominated solutions appear for $V_L \simeq V_{CKM}$ and
$0.01\,{\rm eV} \lesssim m_1 \lesssim 1\,{\rm eV}$ 
\cite{riotto2,strongSO10}, the largest possible $m_1$ values in
$SO(10)$-inspired leptogenesis since tauon-dominated solutions are realised for
$m_1 \lesssim 0.07\,{\rm eV}$ (just at the edge of highest values 
allowed by cosmological observations).  Thus they open a new region in low energy neutrino parameter space though
currently disfavoured by the cosmological observations.  
We have not found electron-dominated solutions
as in the supersymmetric framework \cite{marfatia,susy}.
\end{itemize}
These are the main effects induced by $V_L \simeq V_{CKM}$ that we want to understand analytically 
unpacking the solution we found.

\subsection{$C\!P$ asymmetries flavour ratio}

We have seen that for $V_L = I$
the Eq.~(\ref{ratiosVLI}) immediately shows how in this approximation one cannot 
reproduce electron and muon-dominated solutions.  This result changes
turning on $V_L \simeq V_{CKM}$. 
We can use these two  approximations in the Eq.~(\ref{ve2alAN}) 
\bea\label{approx1}
m^\star_{D \a 2} & = & \sum_{k} V_{L  k \a } \, m_{D k} \, U^{\star}_{R k2}  \simeq 
m_{D2}\,\left(V_{L 2 \a }\,U^{\star}_{R 22} + V_{L 3\a}\, A^{\star}_{32} \right) \,  ,   \\  \nonumber
m_{D\a 3} & = & \sum_{l} V_{L  l \a } \, m_{D l} \, U_{R l3}  \simeq   m_{D3} \, V^{\star}_{L 3\a }\, U_{R 33}   \,   .
\eea
Notice that they give $\ve_{2e}=\ve_{2\mu}\simeq 0$ for $V_{L}=I$, something acceptable
if one wants just to describe solutions giving successful leptogenesis since
 as we have seen, for $V_{L}=I$, there are not electron and muon-dominated solutions since the $C\!P$
asymmetries are too small. 

Using these approximations, from the Eq.~(\ref{ve2alAN}) one obtains for the three $C\!P$ flavour asymmetries
\bea\label{ve2alANbis}
\ve_{2e} & \simeq & {3\,m^2_{D2} \over 16\, \pi \, v^2}\,
{|(\widetilde{m}_{\nu})_{11}| \over m_1 \, m_2 \, m_3}\,
{{\rm Im}[V_{L 12} \, V_{L 13}^{\star} \, U^2_{R 33} \, (A^{\star}_{32})^2]
\over |(\widetilde{m}_{\nu}^{-1})_{33}|^{2} + |(\widetilde{m}_{\nu}^{-1})_{23}|^{2}}   \,   , 
\\ \label{ve2muanA}
\ve_{2\mu}  & \simeq & {3 \,m^2_{D2} \over 16\, \pi \, v^2}\,
{|(\widetilde{m}_{\nu})_{11}| \over m_1 \, m_2 \, m_3}\,
{{\rm Im}[V_{L 22} \, V_{L 23}^{\star} \, U^{\star}_{R22} \,  U^2_{R 33} \, A^{\star}_{32}]
\over |(\widetilde{m}_{\nu}^{-1})_{33}|^{2} + |(\widetilde{m}_{\nu}^{-1})_{23}|^{2}}   \,   ,
\\ \nonumber
\ve_{2\t} & \simeq &  {3 \,m^2_{D2} \over 16\, \pi \, v^2}\,
{|(\widetilde{m}_{\nu})_{11}| \over m_1 \, m_2 \, m_3}\,
{|V_{L 33}|^2 \, {\rm Im}[ U^2_{R 33} \, (A^{\star}_{32})^2]
\over |(\widetilde{m}_{\nu}^{-1})_{33}|^{2} + |(\widetilde{m}_{\nu}^{-1})_{23}|^{2}}   \,    ,
\eea
implying
\be\label{ratiosVL}
\ve_{2\t}^{\rm max}: \ve_{2\m}^{\rm max} : \ve_{2e}^{\rm max} 
\simeq 1 : |V_{L 23}| : |V_{L 21} \,  V_{L 31}| \,  ,
\ee
showing that this time, turning on the mixing angles in $V_L$, the tauon-dominated solutions are still favoured but 
potentially one can also have muon and even electron-dominated solutions. 
\footnote{Notice also that turning on $V_L \neq I$ does not change the result
that dominantly the $\ve_{2\a} \propto m_{D2}^2 = \a_2^2 \,{m_{\rm c}}^2$ while they do not depend on
$\a_1$ and $\a_3$.}

\subsection{Tauon-dominated solutions and strong thermal leptogenesis}

Let us start from the tauon flavour contribution.  As already pointed out, for $V_L = I$ this is the only contribution  that can reproduce the observed asymmetry \cite{riotto1} and, therefore, one has the simplified
result $\left.N_{B-L}^{\rm lep,f} \right|_{V_L = I} \simeq N^{\rm lep,f}_{\D_{\tau}}$. 
A full analytic description was given in \cite{SO10decription}, we already 
reviewed the analytic expressions for $\ve_{2\tau}$ (Eq.~(\ref{ve2tauVLI})), for the
flavour decay parameters $K_{1\t}$, $K_{2\t}$ (see Eqs. (\ref{K1tauVLI}) and (\ref{K2tauVLI})) 
and for the  final asymmetry (Eq.~(\ref{NBmLfVLI})). 

In the left panels Fig.~2 we are plotting the behaviour of all these quantities for a specific choice
of the low energy neutrino parameters: we adopted the best fit values for $\theta_{12}$
and $\theta_{13}$ and then $\theta_{23}= 42^{\circ}$, $\delta = -0.6\,\pi$. As one can see from the
scatter plot in Fig.~1 in the plane $\d$ versus $\theta_{23}$, for this choice of values the observed 
asymmetry cannot be reproduced for $V_L = I$ (light blue points) since $\theta_{23}$ is too large. The plots in the bottom left panel of Fig.~2 confirm the result of the scatter plots. In the panels the thin black lines are the analytic
expressions and one can see that they perfectly reproduce all numerical results. 
\begin{figure}
\hspace*{48mm} $V_L = I$
\hspace*{35mm} $V_L \neq I$
\begin{center}
\psfig{file=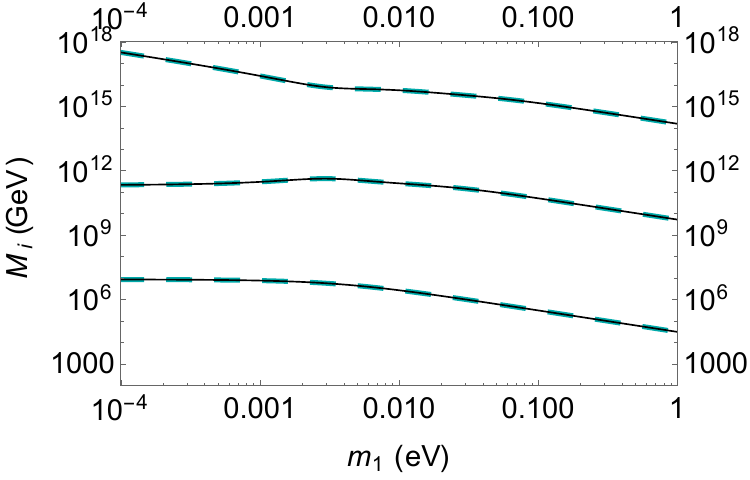,height=37mm,width=45mm}
\hspace{3mm}
\psfig{file=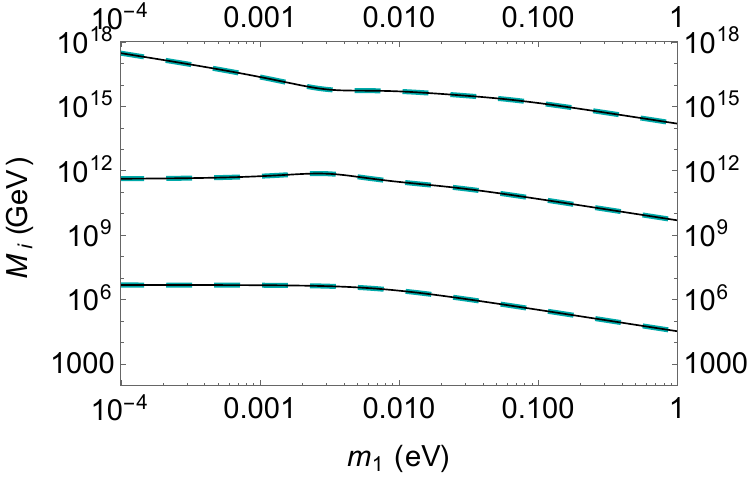,height=37mm,width=45mm}   \\
\psfig{file=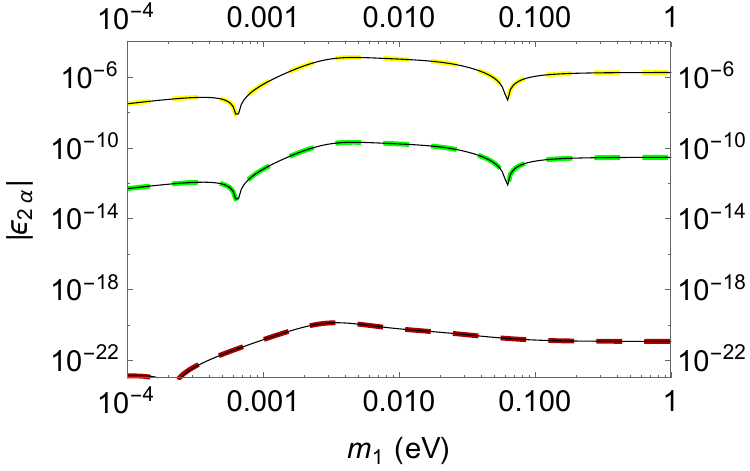,height=37mm,width=45mm}
\hspace{3mm}
\psfig{file=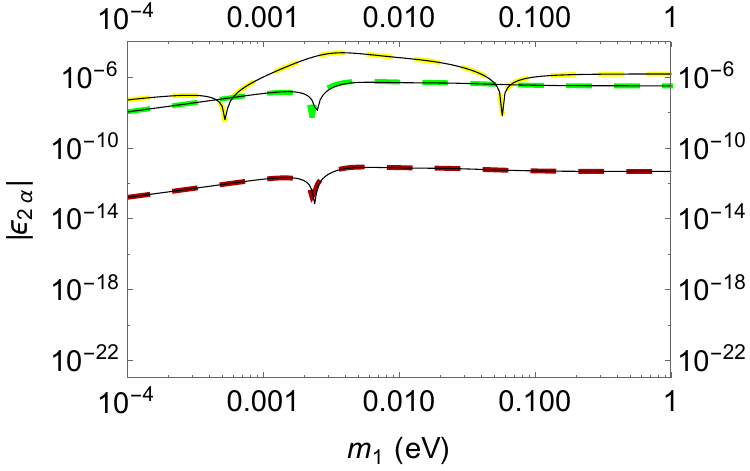,height=37mm,width=45mm} \\
\psfig{file=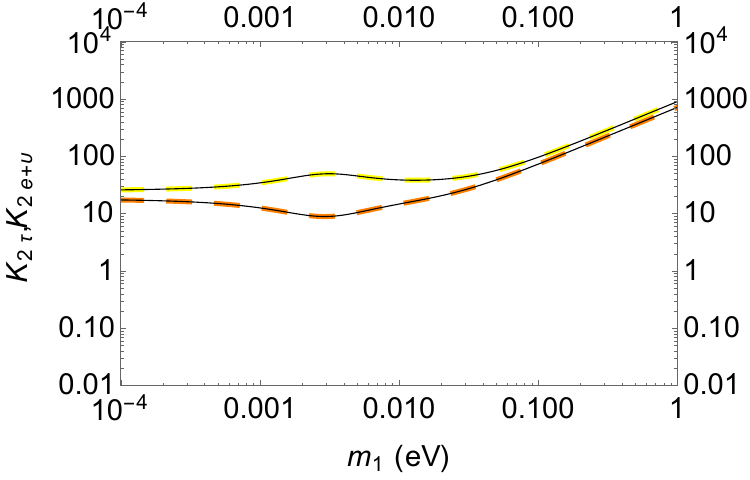,height=37mm,width=45mm}
\hspace{3mm}
\psfig{file=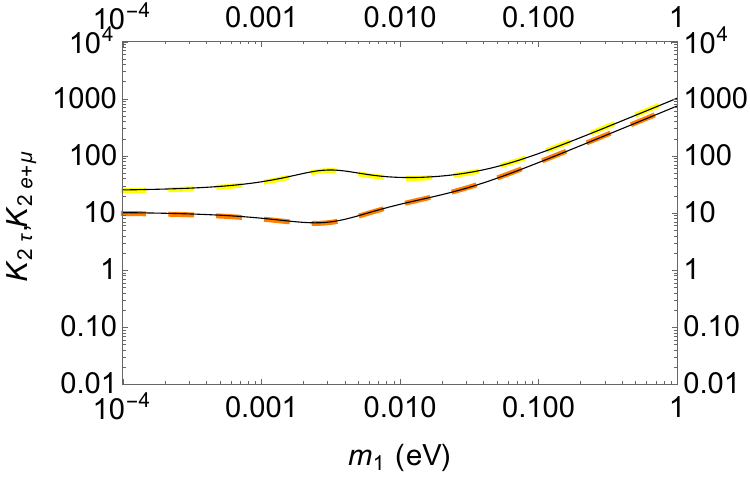,height=37mm,width=45mm}  \\
\psfig{file=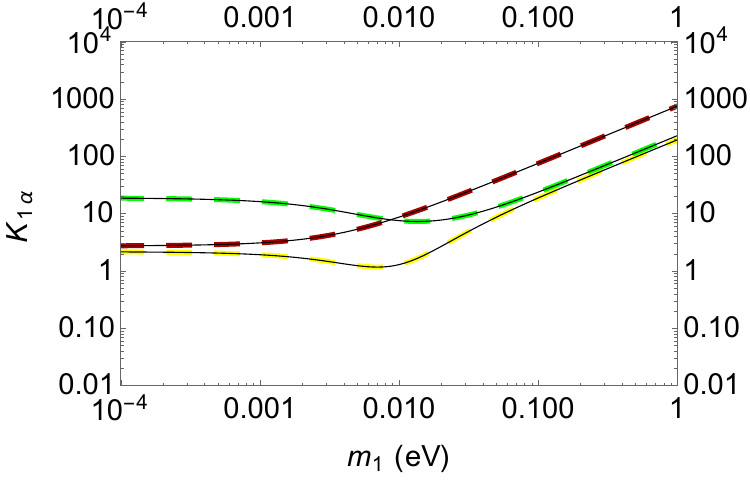,height=37mm,width=45mm}
\hspace{3mm}
\psfig{file=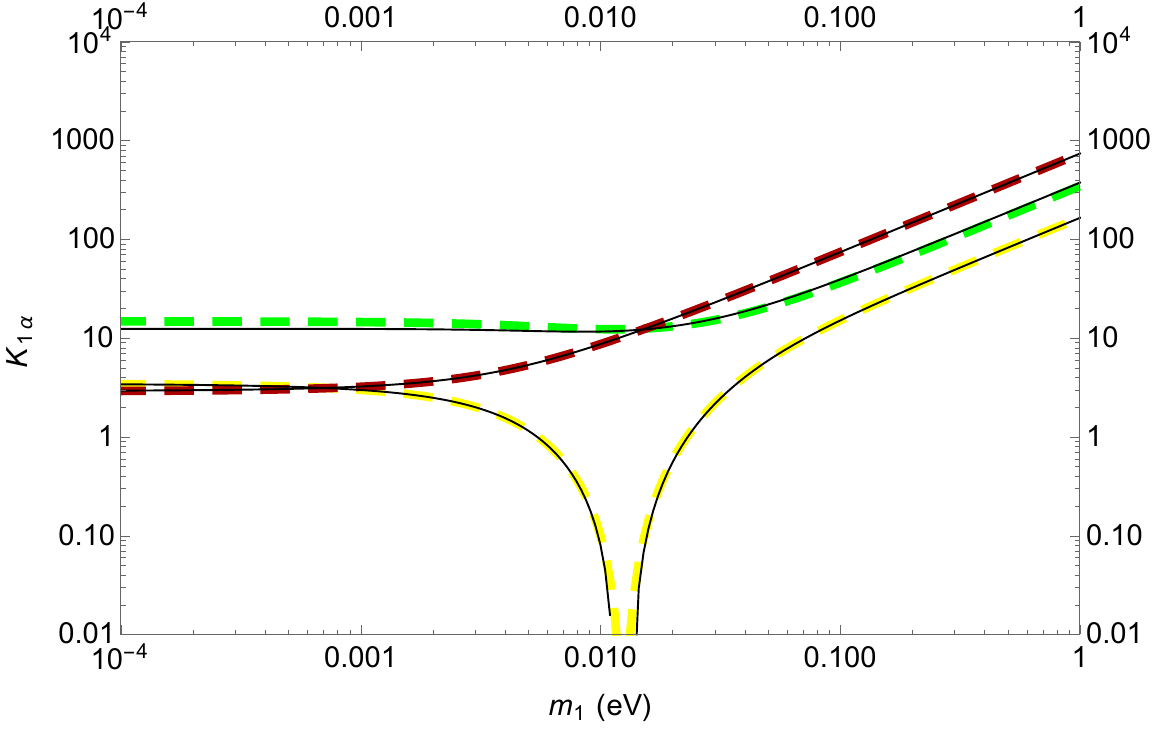,height=37mm,width=45mm}  \\
\psfig{file=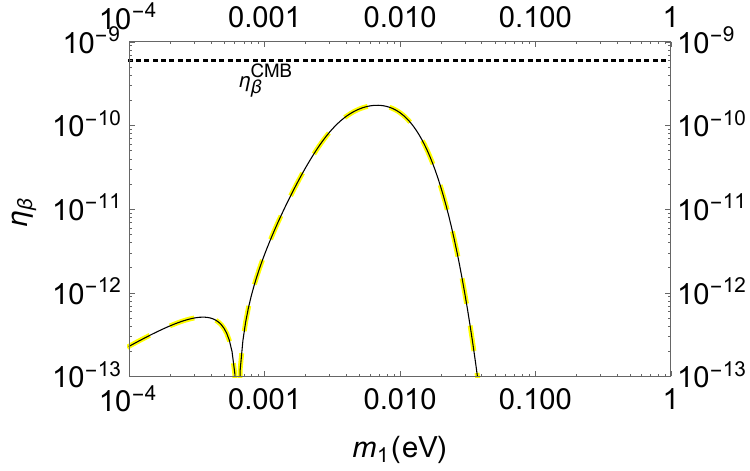,height=37mm,width=45mm}
\hspace{3mm}
\psfig{file=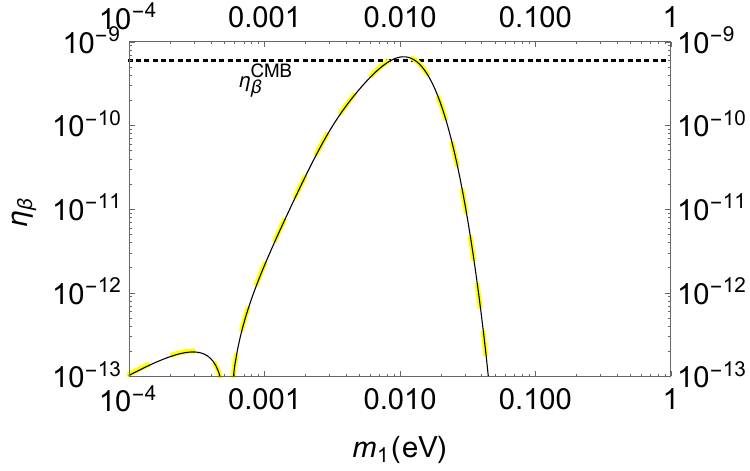,height=37mm,width=45mm}
\end{center}
\vspace{-8mm}
\caption{Example of (tauon-dominated) strong thermal solution.
{\em Left panels}: $V_L = I$, $(\a_1,\a_2,\a_3) = (5,5,5)$,
$(\theta_{13}, \theta_{12}, \theta_{23})=(8.4^{\circ},33^{\circ},42^{\circ})$, 
$(\d,\r,\s)=(-0.6\pi, 0.23 \pi, 0.78\,\pi)$;
{\em Right panels}: Same as for left panels but $V_L \neq I$ and 
$(\theta^L_{13}, \theta^L_{12}, \theta^L_{23})=(0.1^{\circ},9.5^{\circ},2.4^{\circ})$ and
$(\d_L,\r_L,\s_L)=(1.2\pi, 0.02 \pi, 1.15 \,\pi)$.  
All thin black lines are the analytical expressions for each corresponding quantity. 
The long-dashed coloured lines indicate the numerical results (same colour
code as in Fig.~1: yellow for tauon flavour, green for  muon flavour and red for electron flavour, 
the orange lines refer to the e+$\mu$ flavour).}
\end{figure}

For $V_L \simeq V_{CKM}$ all the expressions get generalised in the way we have seen. Let us 
specialise them and make more explicit for the tauon-dominated case. 
First of all for the tauonic $C\!P$ asymmetry,
considering only the dominant terms in the Eq.~(\ref{ve2alAN}) we find 
\be\label{ve2alANtau}
\ve_{2\tau}  \simeq  {3\,m^2_{D2} \over 16\, \pi \, v^2}\,
{|(\widetilde{m}_{\nu})_{11}| \over m_1 \, m_2 \, m_3}\,
{|(\widetilde{m}_{\nu}^{-1})_{23}| \over |(\widetilde{m}_{\nu}^{-1})_{33}|}\,
{[|V_{L33}|^2 \, (|(\widetilde{m}_{\nu}^{-1})_{23}|/|(\widetilde{m}_{\nu}^{-1})_{33}|)
\sin \a_{L}^{\tau A} + |V_{L33}|\,|V_{L23}|\, \sin \a_{L}^{\tau B} ]
\over |(\widetilde{m}_{\nu}^{-1})_{33}|^{2} + |(\widetilde{m}_{\nu}^{-1})_{23}|^{2}} \,  ,
\ee
where
\bea
\a_L^{\tau A} & = & 
{\rm Arg}\left[\widetilde{m}_{\nu 11}\right]  - 2\,{\rm Arg}[(\widetilde{m}^{-1}_{\nu})_{23}] 
- \pi -2\,(\rho+\s) -2\,(\rho_L+\s_L)  \,  ,
\\
\a_L^{\tau B} & = & 
{\rm Arg}\left[\widetilde{m}_{\nu 11} \right]  - \,{\rm Arg}[(\widetilde{m}^{-1}_{\nu})_{23}] 
- {\rm Arg}[(\widetilde{m}^{-1}_{\nu})_{33}]   -2\,(\rho+\s) -2\,(\rho_L+\s_L)  \,  ,
\eea
that generalises the Eq.~(\ref{ve2muanA}) for $V_L \neq I$. Notice that the second term is subdominant
but still gives an important correction if $\theta_{23}^L$ is not too small. This analytic expression 
produced the black thin line in the second right panel in Fig.~2 and one can see that it perfectly 
fits the numerical result. 

For the flavour decay parameters $K_{1\t}$ and $K_{2\t}$ we find respectively
\be\label{K1tau}
K_{1\t} \simeq 
{1\over m_{\star}}\,
\left( {|\widetilde{m}_{\nu 13}|^2 \over |\widetilde{m}_{\nu 11}|}\,|V_{L33}|^2
+2\,{V_{L23}\,V_{L33}^{\star} \over |\widetilde{m}_{\nu 11}|}\,
{\rm Re}\left[\widetilde{m}_{\nu 12}^\star \, \widetilde{m}_{\nu 13} \right]
+|V_{L 23}|^2 \, {|\widetilde{m}_{\nu 13}|^2 \over |\widetilde{m}_{\nu 11}|} \right)
\ee
and 
\be
K_{2\t} \simeq
{m_1\,m_2\,m_3 \over m_{\star}}\, 
{|(\widetilde{m}_{\nu}^{-1})_{2 3}|^2 \over |\widetilde{m}_{\nu 11}|\, |(\widetilde{m}_{\nu}^{-1})_{33}|} \,  .
\ee
These analytic expressions also very well agree with the numerical results as it can be seen in the example
of Fig.~2 in the right panels. In the case of $K_{1\t}$ one needs more accuracy than for $K_{2\t}$
since it suppresses exponentially the asymmetry and one needs to add also terms 
$\propto V_{L23}$ in order to get correctly the tauonic contribution to $\eta_B$, 
as one can see in the last right panel of Fig.~2 where the analytic contribution (thin black line) nicely matches
the numerical results (yellow dashed line).  There one can notice how the final asymmetry gets enhanced
\footnote{The reason why the peak of the asymmetry is just above the observed value is because we
have deliberately chosen a solution at the border of the allowed region.}
by almost two orders of magnitude compared to the case $V_L=I$ and the main reason is that
turning on $V_L \simeq V_{CKM}$ makes now possible to have $K_{1\t} \ll 1$ at larger values
of $\theta_{23}$ and smaller values of $\d$ something quite important considering that
long baseline experiments such as NO$\nu$A and T2K are right now testing these parameters
and in particular the deviation of $\theta_{23}$ from maximal mixing.

It is quite straightforward to extend the derivation of the upper bound on $\theta_{23}$
presented in \cite{SO10decription} for $V_L = I$ turning on $V_L \simeq V_{CKM}$, finding
\be
\theta_{23} \lesssim {\rm arctan} \left[{m_{\rm atm}\,s_{13}/\sqrt{2} \over 
(m_1 +m_{\rm sol})\, c_{13}\,c_{12}\,s_{12} -V_{L12}\,(m_{\rm atm}-m_{\rm sol}- s^2_{12}\,m_1)/\sqrt{2}} \right] \sim 45^{\circ} \,  ,
\ee
where we took into account that $2\s-\d \simeq -\pi/4$ and this yields the factor $1/\sqrt{2}$ in the numerator.
In this case the largest angle $\theta_{12}^L$ gives the dominant effect. 

We can also easily understand why the lower bound on $m_{ee}$ gets strongly relaxed from $m_{ee} \gtrsim 1\,{\rm meV}$ to $m_{ee} \gtrsim 0.1\,{\rm meV}$ considering that
\be
|\widetilde{m}_{\nu 11}| \simeq \left|\cos^2 \theta_{12}^L\, m_{\nu ee} e^{i\,\rho_L} +{1\over 2}
\, \sin 2\theta_{12}^L \, m_{\nu e \mu}\right| \,   .
\ee
 The lower bound $|m_{ee}| \gtrsim 1\,{\rm meV}$ that was holding for $V_L = I$ translates now,
 for $V_L \simeq V_{CKM}$, into $|\widetilde{m}_{\nu 11}| \gtrsim 1\,{\rm meV}$. 
 \footnote{This lower bound can be understood considering that 
 $K_{1\tau} \propto |\widetilde{m}_{\nu 11}|^{-1}$ (see Eq.~(\ref{K1tau})).}
 It is then possible to have  the second term in $|\widetilde{m}_{\nu 11}|$ dominating
 and saturating the lower bound while $m_{\nu ee} \ll 1\,{\rm meV}$. However, a lower bound still exists
 and $m_{ee}$  cannot be arbitrary small. It is interesting actually to see from the panel in Fig.~1 
 showing $m_{ee}$ versus $\d$, that there seems to be values of $\d$ for which the lower bound becomes more stringent
 and that in any case the bulk of points is well above $1\,{\rm meV}$ and within reach of future experiments. This is   an interesting feature of $SO(10)$-inspired leptogenesis. For the strong thermal points of course this 
is true even more stringently, since in this case $m_{ee} \gtrsim 10\,{\rm meV}$ \cite{strongSO10} and a signal should be  in the reach of future experiments despite the fact that neutrino masses are NO.

We also want to remind, in conclusion of this subsection, that tauon-dominated solutions do not imply any fine-tuning in the seesaw formula, indeed the orthogonal matrix for these solutions has all entries $|\O_{ij}|\lesssim 1$, also for this reason they have  certainly to be regarded as the canonical and most attractive solutions.

\subsection{Muon-dominated solutions}

For $V_L =I$ there are no muon-dominated solutions \cite{riotto2}. In the {\em left panels} of Fig.~3 we show the dependence of different quantities on $m_1$
for $V_L =I$, $(\a_1,\a_2,\a_3)=(5,5,5)$ and for the indicated set of values 
of the low energy parameters. In particular one can notice 
how the $C\!P$ flavoured asymmetries (second left panel from top) respect the strong hierarchical pattern 
in Eq.~(\ref{ratiosVLI}) and even though both the wash-out at the production and, more importantly,
from lightest RH neutrinos are negligible in the muon flavour, the final asymmetry (last left panel)
falls many orders of magnitude below the observed value. 
\begin{figure}
\hspace*{48mm} $V_L = I$
\hspace*{35mm} $V_L \neq I$
\begin{center}
\psfig{file=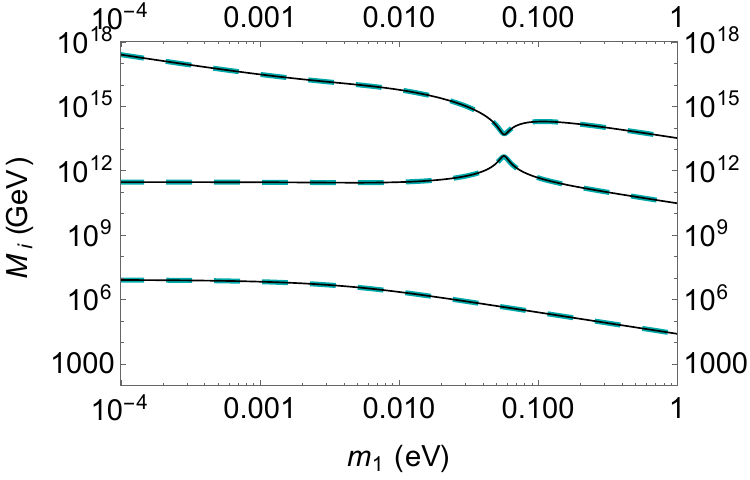,height=37mm,width=45mm}
\hspace{3mm}
\psfig{file=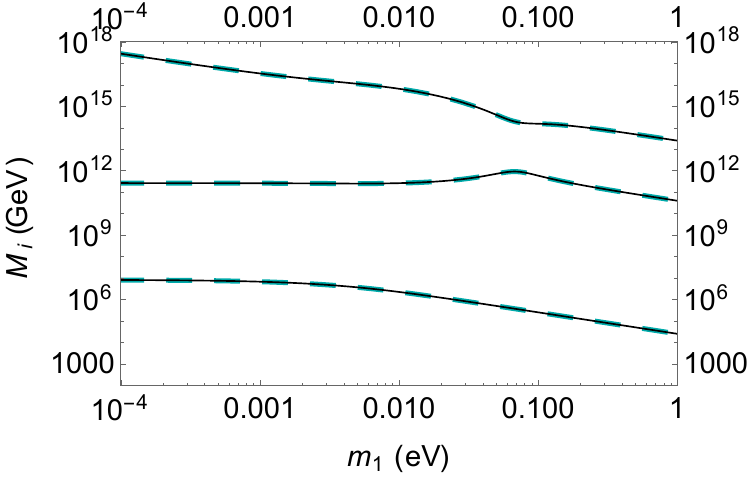,height=37mm,width=45mm}   \\
\psfig{file=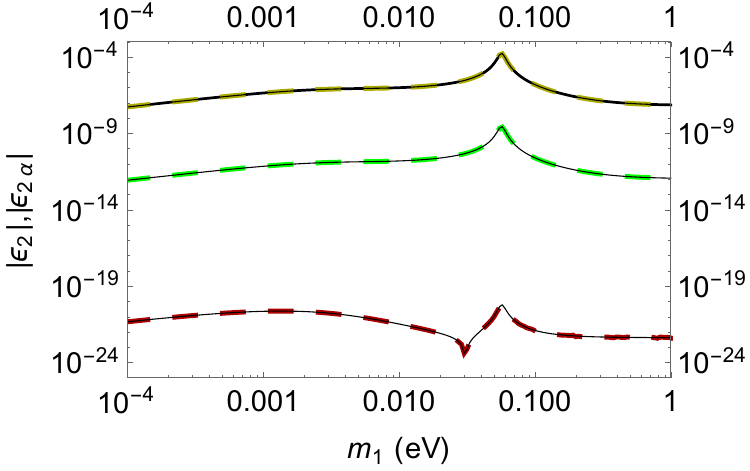,height=37mm,width=45mm}
\hspace{3mm}
\psfig{file=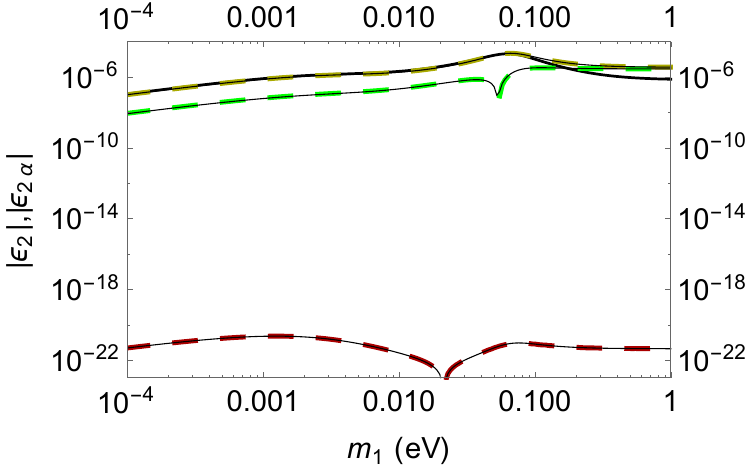,height=37mm,width=45mm} \\
\psfig{file=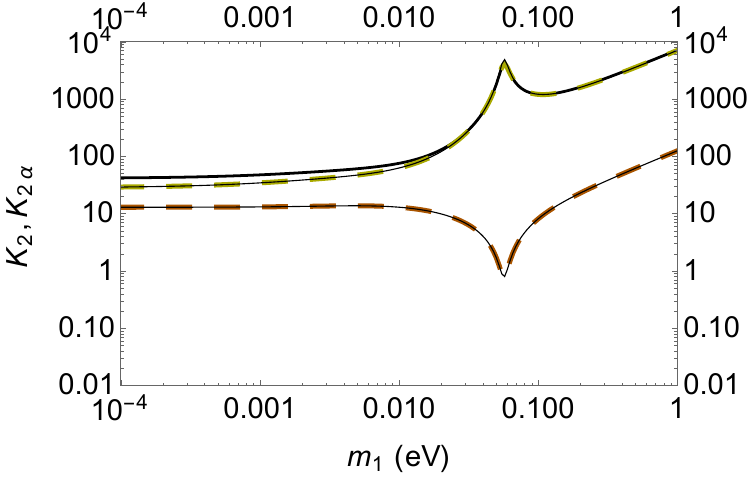,height=37mm,width=45mm}
\hspace{3mm}
\psfig{file=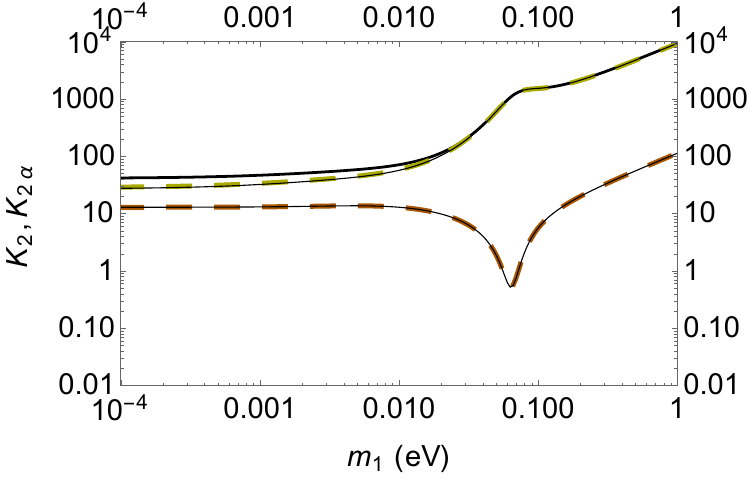,height=37mm,width=45mm}  \\
\psfig{file=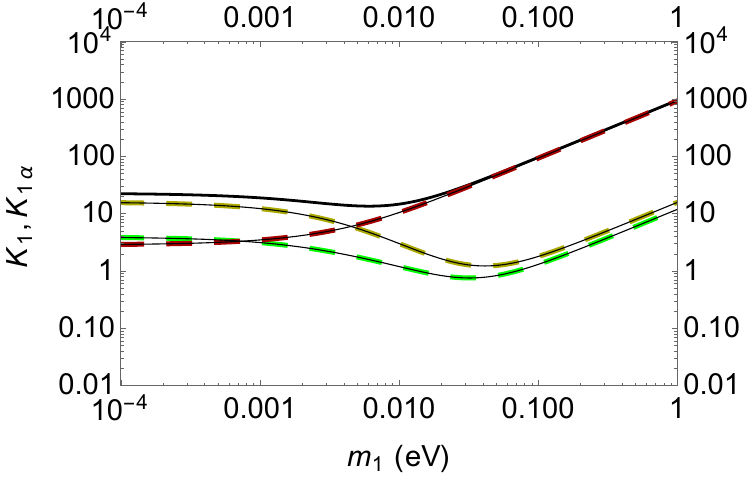,height=37mm,width=45mm}
\hspace{3mm}
\psfig{file=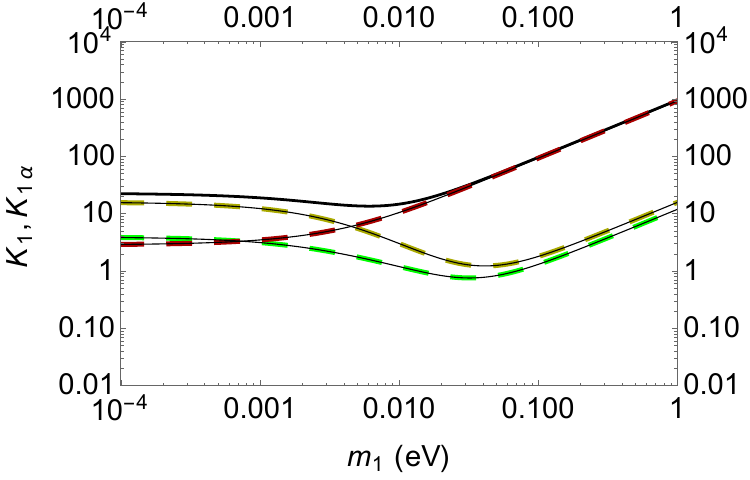,height=37mm,width=45mm}  \\
\psfig{file=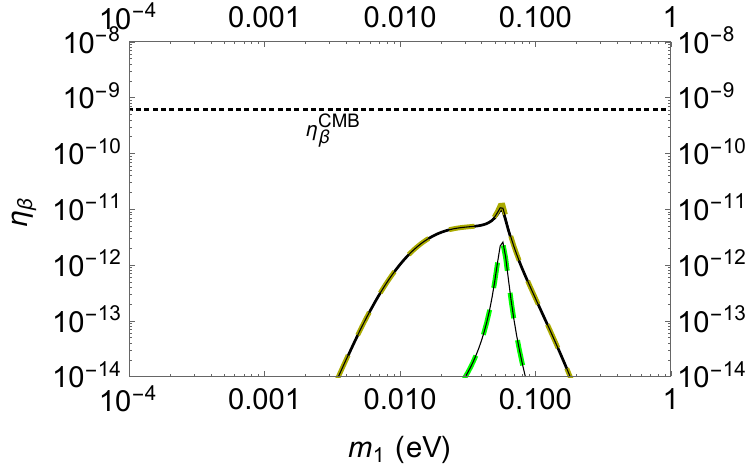,height=37mm,width=45mm}
\hspace{3mm}
\psfig{file=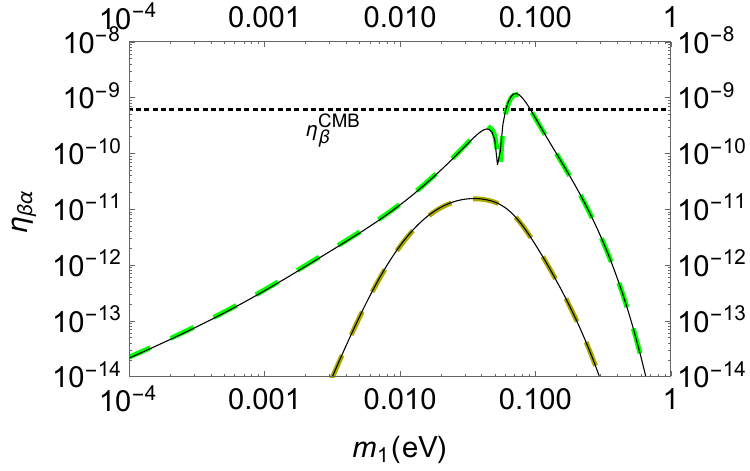,height=37mm,width=45mm}
\end{center}
\vspace{-8mm}
\caption{Example of muon-dominated solution.
{\em Left panels}: $V_L = I$, $(\a_1,\a_2,\a_3) = (5,5,5)$,
$(\theta_{13}, \theta_{12}, \theta_{23})=(8.4^{\circ},33^{\circ},41^{\circ})$, 
$(\d,\r,\s)=(-0.3\pi, 0, 0.5\,\pi)$;
{\em Right panels}: same as for left panels except that $\theta_{23}^L = 2.4^{\circ}$ and
$\sigma_L =-1.7\pi$ (the values of $\delta_L$ and $\rho_L$ are irrelevant, since they
cancel out for $\theta^L_{12}=\theta^L_{13}=0$).  
All thin black lines are the analytical expressions for each quantity. The long-dashed coloured lines indicate the numerical results (same colour
code as in Fig.~1: yellow for tauon flavour, green for  muon flavour and red for electron flavour, 
the orange lines refer to the e+$\mu$ flavour).}
\end{figure}
Turning on $V_L \simeq V_{CKM}$, as we have seen in the scatter plots of Fig.~1, one does obtain
muon-dominated solutions.  We want to show here analytically how this occurs and derive an analytic expression
that reproduces correctly the muon asymmetry. 

First of all let us notice that the result in Eq.~(\ref{ratiosVL}) is well illustrated by the right panels of Fig.~3. 
They are obtained for the same set of values as in the left panels except that now $V_L \neq I$,
with the only non-vanishing angle $\theta_{23}^L = 2.4^{\circ} \simeq \theta_{23}^{CKM}$
and also non-vanishing values of the phases $\sigma_L$ and $\rho_L$. One can see that the muon asymmetry
now gets enhanced compared to the right panel where $\theta_{23}^L = 0$ while the electron asymmetry
is unchanged. This is in complete agreement with the result in Eq.~(\ref{ratiosVL}): for muon-dominated
solutions it is then crucial to have non-vanishing $\theta_{23}^L$.

The Eq.~(\ref{ve2muanA}) neglects a term $\propto |V_{L 32}|^2$ 
that for a more accurate result we now need to add.
Going back to the Eq.~(\ref{ve2alAN}), similarly to the 
tauon asymmetry, there are two terms $\propto m_{D2}^2$ and one obtains 
\bea
\ve_{2\mu} \simeq \ve_{2\mu}^{V_L}  & = &  {3 \,m^2_{D2} \over 16\, \pi \, v^2}\,
{|(\widetilde{m}_{\nu})_{11}| \over m_1 \, m_2 \, m_3}\,   \\ \nonumber
& \times  &  {|(\widetilde{m}_{\nu}^{-1})_{23}| \over |(\widetilde{m}_{\nu}^{-1})_{33}|} \, 
 {|V_{L22}| \, |V_{L 32}| \, \sin\a_{L}^{\m A} + 
 |V_{L 32}|^2 \, (|(\widetilde{m}_{\nu}^{-1})_{23}| / |(\widetilde{m}_{\nu}^{-1})_{33}| )\sin\a_{L}^{\m B}\over |(\widetilde{m}_{\nu}^{-1})_{33}|^{2} + |(\widetilde{m}_{\nu}^{-1})_{23}|^{2}}  \,   ,
\eea 
where 
\bea
\a_L^{\m A} & = & 
{\rm Arg}\left[\widetilde{m}_{\nu 11}\right]  - {\rm Arg}[(\widetilde{m}^{-1}_{\nu})_{23}] 
- {\rm Arg}[(\widetilde{m}^{-1}_{\nu})_{33}]  -2\,(\rho+\s) -2\,(\rho_L+\s_L)  \,  ,
\\
\a_L^{\m B} & = & 
{\rm Arg}\left[\widetilde{m}_{\nu 11} \right]  - 2\,{\rm Arg}[(\widetilde{m}^{-1}_{\nu})_{23}] 
- \pi -2\,(\rho+\s) -2\,(\rho_L+\s_L)  \,   .
\eea
This analytic expression for $\ve_{2\mu}$ 
perfectly matches the numerical result  in Fig.~2 (respectively the thin black line and the dashed green line). 
For completeness we also fit the muonic asymmetry  for very small or vanishing $\theta_{23}^L$, 
adding the following term ($\propto m_{D2}^4/m_{D3}^2$)
\be
\ve_{2\m}^I \simeq  {3 \,m^2_{D2} \over 16\, \pi \, v^2}\,{m_{D2}^2 \over m_{D3}^2}\, 
 {|\widetilde{m}_{\nu 11}| \over m_1 \, m_2 \, m_3}\, 
 {|(\widetilde{m}_{\nu}^{-1})_{23}|^2 \over |(\widetilde{m}_{\nu}^{-1})_{33}|^2}\,
 { |V_{L 22}|^2 \over |(\widetilde{m}_{\nu}^{-1})_{33}|^{2} + |(\widetilde{m}_{\nu}^{-1})_{23}|^{2}}
 \, \sin \widetilde{\a}_L \,   ,
\ee
where
\be
\widetilde{\a}_L  =  
{\rm Arg}\left[\widetilde{m}_{\nu 11}\right]  - 2\, {\rm Arg}[(\widetilde{m}^{-1}_{\nu})_{23}] 
-2\,(\rho+\s) -2\,(\rho_L+\s_L)  \,   ,
\ee
so that $\ve_{2\mu} \simeq \ve_{2\m}^I + \ve_{2\mu}^{V_L}$.
Like in the limit $V_L \rightarrow I$, 
this term is not sufficiently large to reproduce the observed baryon asymmetry 
(not at least for $\a_2 \lesssim 5$), however by adding this term we could also reproduce $\ve_{2\m}$ in the case shown in  Fig.~2, for a tauon-dominated solution. 

One can also understand why muon-dominated solutions exist only in the range 
$0.01 \, {\rm eV} \lesssim m_1 \lesssim 1\,{\rm eV}$ from the expressions 
Eqs.~(\ref{KialVL}) specialised for $K_{1\m}$ and $K_{2\m}$. 
In this case the deviations from $V_L =I$
give only corrections, as we will show explicitly, and we can first consider the simplified
expressions for $V_L =I$. First of all we can write
\bea
K_{1\m} & = & {m^2_{D2}\,|U_{R 21}|^2 \over m_{\star}\,M_1} = {|m_{\nu e \mu}|^2 \over m_{\star}\,m_{ee}} \\ \nonumber
             & =&  {c^2_{13}\,\left|c_{12}\,s_{12}\,c_{23}\,(m_2-m_1\,e^{2i\rho}) - 
             s_{13}\, s_{23}\,e^{2 i \s} \left[e^{i\d}(m_1\,c^2_{12}+m_2\, s^2_{12}) 
             -m_3 \, e^{-i\d}\right]\right|^2 \over m_{\star} \, 
             |m_1 \, c_{12}^2\,c_{13}^2 \,e^{2i\rho} + m_2\,s^2_{12}\,c^2_{13} +m_3\,s^2_{13}\, 
             e^{i\,(2\s - \d)}|} \,  .
\eea
From this general expression one can easily see that in the hierarchical limit $m_1 \lesssim m_{\rm sol}$
one has
\be
K_{1\m} \rightarrow c^2_{12}\,c_{23}^2\,{m_{\rm sol} \over m_{\star}} \simeq 3 \,  ,
\ee
giving a too strong suppression  in the hierarchical limit. 
On the other hand for $m_1 \gtrsim m_{\rm sol}$
one has $m_1 \simeq m_2$ and the dominant term in the numerator cancels out for 
$\rho \simeq n\,\pi$ and one can have $K_{1\m} \lesssim 1$. However,
 for $m_1 \gtrsim m_{\rm atm}$, in the quasi-degenerate climit, one has
 \be
 K_{1\m} \rightarrow s^2_{23}\,s^2_{13}\,{m_1 \over m_{\star}}\,
 \left|1-e^{2i(\s -\d)} \right|^2 \,  .
 \ee
This can be still made small or even vanishing (for $\d = \s$). 
The wash-out at the production is described by $K_{2\m}$ (since $K_{2e} \lll K_{2\m}$)
given by
\bea
K_{2\m} & = & 
{m^2_{D2} \over m_{\star}\,M_2} = 
{m_1 \, m_2 \, m_3 |(m_{\nu}^{-1})_{\t\t}| \over m_{\star}\,m_{ee}} \,  ,
\eea
that in the quasi-degenerate limit becomes
\be
K_{2\m} \rightarrow {m_1 \over m_{\star}}\,|s^2_{23}+c^2_{23}\,c^2_{13}\,e^{-2i\s}| \,  .
\ee
One can have a cancellation around $\s= (2m+1)\,\pi/2$ but away from this condition, for large values of $m_1$,
$K_{2\m}$ increases linearly with $m_1$. The larger is $m_1$, the sharper the 
conditions $\rho \simeq n\,\pi$ and $\d=\s= (2k+1)\,\pi/2$ have to be satisfied.
This can be clearly seen in the scatter plots of Fig.~1 (green points),
while the linear increase of $K_{2\m}$ with $m_1$ can be clearly seen in the example shown Fig.~2. 
However, the phase $\a_L \ra 4 \s$ in the quasi-degenerate limit and this leads to an upper bound 
$m_1 \lesssim 1\,{\rm eV}$, that on the other hand is quite relaxed compared to the corresponding one
holding for tauon-dominated solutions discussed in detail in \cite{SO10decription}. 

In the range $0.01\,{\rm eV} \simeq m_{\rm sol} \lesssim m_1 \lesssim 1\,{\rm eV}$
one can have a strong reduction of $K_{2\m}$ and $K_{1\m} \lesssim 1$ and at the same time a sizeable $C\!P$
asymmetry and this explains why in this range there are muon-dominated solutions that, however, are now quite constrained
by the current cosmological upper bound on $m_1$ and also by the upper bound on $m_{ee}$ from $0\nu\b\b$ experiments. 

In order to reproduce accurately the numerical results on the $K_{i\a}$'s vs. $m_1$ 
shown in the Fig.~2, one has to take into account
corrections from $V_L \simeq V_{CKM}$, especially in the case of $K_{2\mu}$. We can first specialise the general expression 
Eq.~(\ref{KialVL})  writing
\be
K_{2\m} = 
{m_1 \, m_2 \, m_3 |(\widetilde{m}_{\nu}^{-1})_{33}| \over m_{\star}\,\widetilde{m}_{\nu 11}}
\,\sum_{k,l}\,V_{L k\m} \, V_{L l \m}^{\star} \, A^{\star}_{k 2}\,  A_{ l 2}  \,  ,
\ee
and then we arrive to the approximate expression
\bea \nonumber
K_{2\m} & \simeq & 
{m_1 \, m_2 \, m_3 |(\widetilde{m}_{\nu}^{-1})_{33}| \over m_{\star}\,\widetilde{m}_{\nu 11}}
\, \times \\
& \times & \left(
|V_{L 22}|^2 + |V_{L 12}|^2 \, {|\widetilde{m}_{\nu 12}|^2 \over |\widetilde{m}_{\nu 11}|^2}
+ 2\,s^L_{23}\,{\rm Re}\left[ 
{(\widetilde{m}_{\nu}^{-1})_{23} \over (\widetilde{m}^{-1}_{\nu})_{33}}\right] 
+ |V_{L 32}|^2 \,  {|(\widetilde{m}_{\nu}^{-1})_{23}|^2 \over |(\widetilde{m}^{-1}_{\nu})_{33}|^2} \right) \,  ,
\eea
that perfectly reproduces (thin black line) the numerical result (dashed orange line) both in Fig.~2 
and in Fig.~3. We also derived an analogous expression for $K_{1\m}$ also reproducing the 
numerical results in Fig.~2 and in Fig.~3.

There is another important aspect to be reported of muon-dominated solutions: they necessarily rely on some amount of fine tuning
in the seesaw formula as it can be understood from the expression of the orthogonal matrix Eq.~(\ref{Omegaapp}). These solutions exist for values of the parameters about  the crossing level solution where $M_2 = M_3$. 
Even though one still has $M_3 \gg M_2$, the value of $M_2$ gets enhanced
and correspondingly the value of $\ve_{2\m}$, this is clearly visible in the panels of Fig.~3. This possibility
relies on the value of $(\widetilde{m}_{\nu}^{-1})_{33}$ in the denominator of $M_2$ to get reduced thanks to
some mild phase cancellation. By itself this is not a problem,  however for
small values of $(\widetilde{m}_{\nu}^{-1})_{33}$ the second and third column in the orthogonal
matrix get correspondingly enhanced, as it can be seen from Eq.~(\ref{Omegaapp}) and this necessarily implies
a fine tuning in the seesaw formula at the level of $1/|\Omega|^2_{ij}$.
These analytical considerations are fully confirmed by the scatter plots obtained numerically. In
Fig.~4 in the left panel we compare scatter plots of solutions in the plane $\theta_{23}$ vs. $m_1$
 imposing the condition $|\Omega_{ij}|^2 < 3, 10, 100$ from left to right: one can notice how
in the first case all muon-dominated solutions, the green points at values $m_1 \gtrsim 0.01\,{\rm eV}$, completely disappear. From this point of view it should be clear that tauon-dominated solutions, the bulk of 
 solutions within $SO(10)$-inspired leptogenesis, are the only completely untuned solutions.
\footnote{Indeed constraints in Fig.~4
do not change increasing $M_{\O}$ for tauon-dominated solutions (yellow and orange points)}
\begin{figure}
\begin{center}
\psfig{file=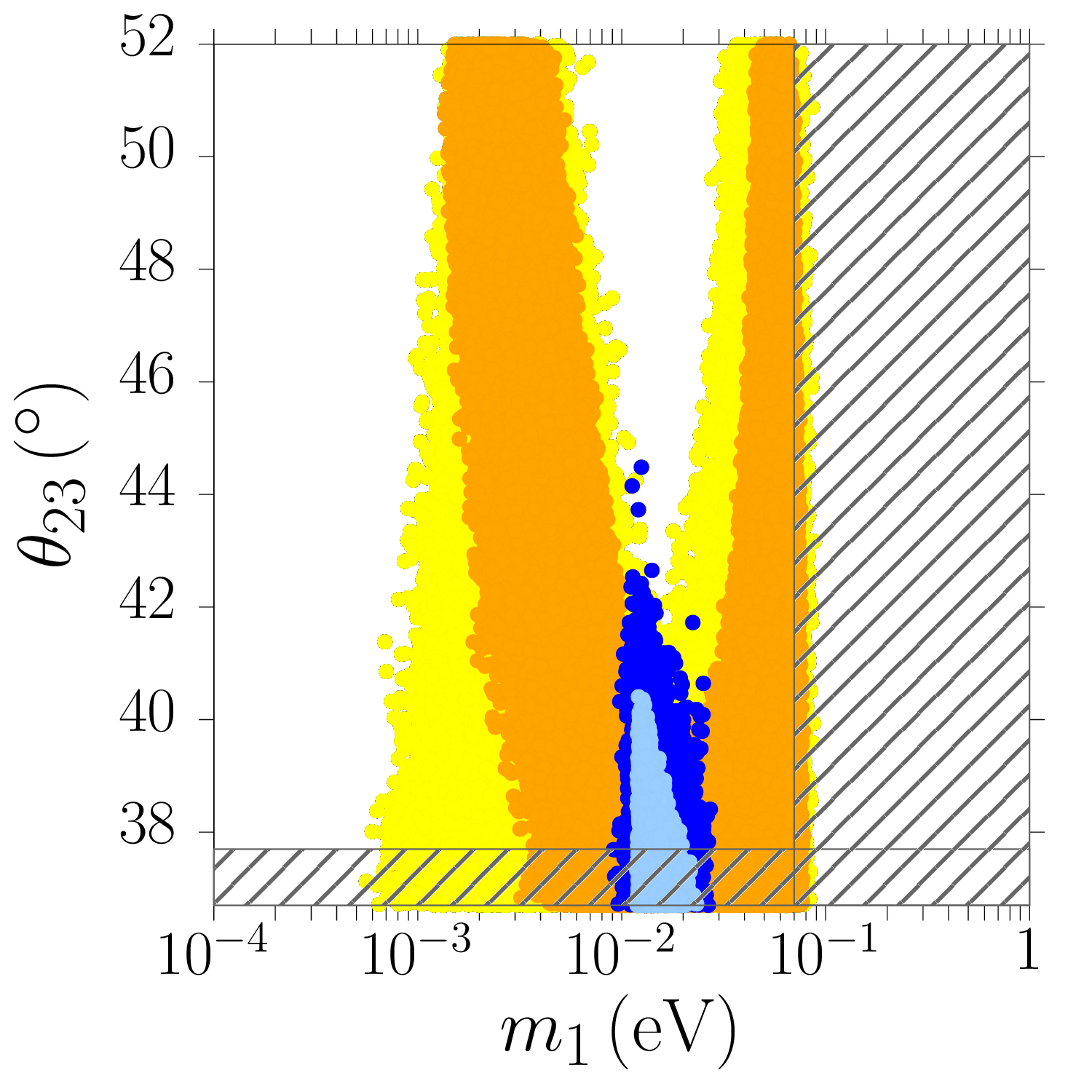,height=50mm,width=52mm}
\hspace{-1mm}
\psfig{file=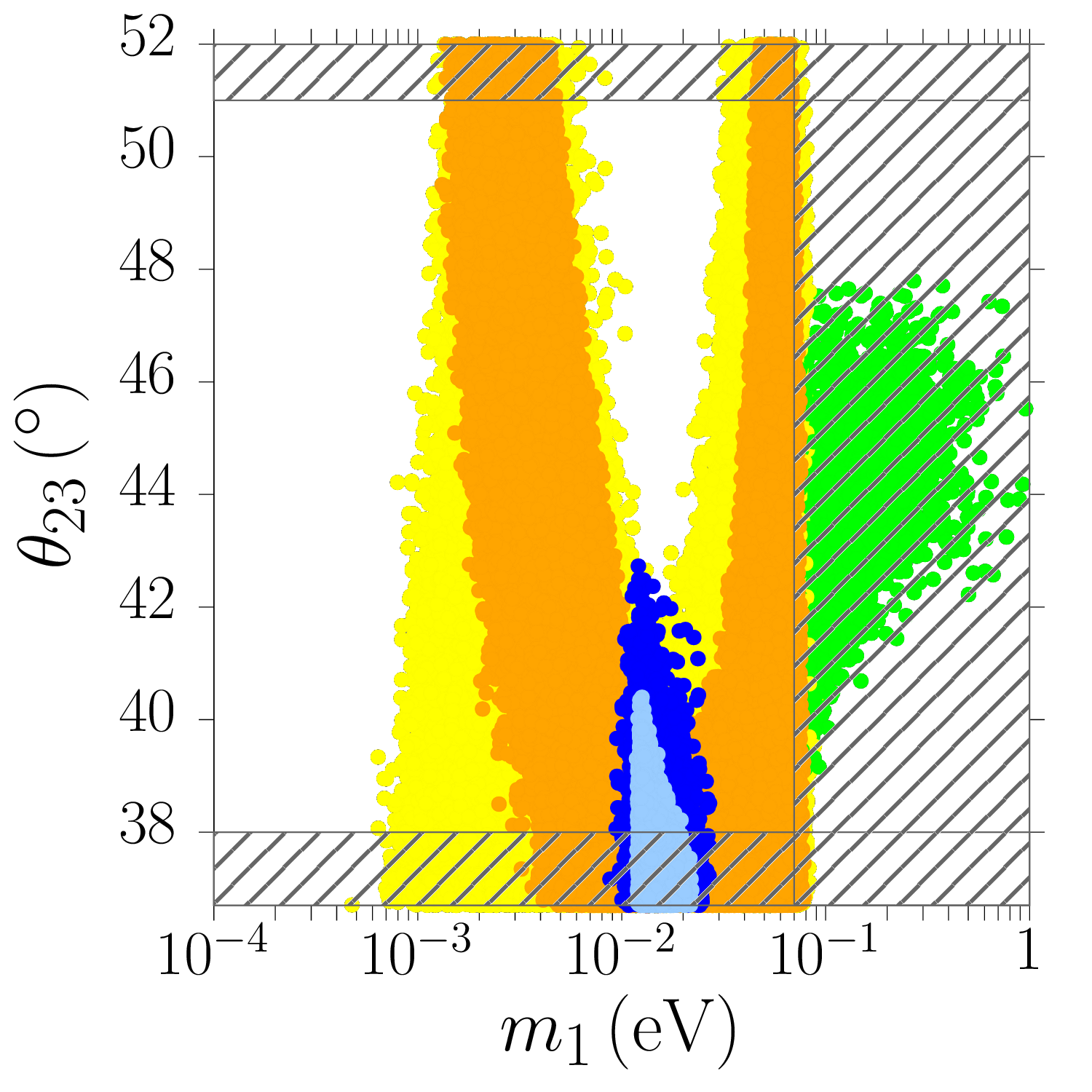,height=50mm,width=52mm}
\hspace{-1mm}
\psfig{file=t23_vs_m1.png,height=50mm,width=52mm} 
\end{center}
\vspace{-5mm}
\caption{Scatter plots in the plane $\theta_{23}$ versus $m_1$ 
as in Fig.~1  but for $M_{\Omega}=3,10,100$ from left to right.
Same colour code as in Fig.~1.}
\label{comparisonMOM}
\end{figure}

\subsection{Electron-dominated solutions?}

In the scatter plots shown in Fig.~1, for $M_{\O}<100$ and $M_3 \geq 2\,M_2$, 
we could not find any electron-dominated solution. 
\footnote{The maximum asymmetry that an electron dominated solution can produce is 
$\eta_B \simeq 3\times 10^{-10}$ for values $m_1 \simeq 4\,{\rm meV}$ corresponding to have
$(\widetilde{m}_{\nu}^{-1})_{33}$ very small, in the vicinity of the crossing level $M_2 \simeq M_3$.
These solutions involve, as stressed already a few times, a high fine tuning in the seesaw formula. The value
of $M_2$ is necessarily capped below $10^{12}\,{\rm GeV}$ since if it goes above, though the  $C\!P$ asymmetry
would grow, the wash-out at the production would occur in the unflavoured regime experiencing a very strong wash-out due to a huge value of $K_{2\tau} \simeq K_2$. In the supersymmetric case the double
value of the $C\!P$ asymmetries and the fact that the transition to the unflavoured regime occurs at 
higher values $M_2 \simeq 10^{12}\,{\rm GeV}\,(1+\tan^2 \beta)$, conspire in a way that sparse 
(very fine-tuned) electronic solutions do appear.}
From the results in Eq.~(\ref{ratiosVLI}) and Eq.~(\ref{ratiosVL}) we can understand the reason: for $V_L=I$ the electron $C\!P$ asymmetry is suppressed
by more than 15 orders of magnitude  compared to the tauonic $C\!P$ asymmetry and even turning on
$V_L \simeq V_{CKM}$ is not enough since $\ve_{2e}$ is still suppressed 
$\propto \theta_{13}^L\,\theta_{12}^L$ compared to the tauonic $C\!P$ asymmetry $\ve_{2\tau}$.

It is however still worth to give briefly analytic expressions for the three quantities 
($\ve_{2e}$, $K_{1e}$, $K_{2e}$)  involved in the calculation of the 
electronic contribution to the asymmetry. The reason is that they can be easily  extended to models 
or frameworks where there might be some enhancement. For example either in $SO(10)$-inspired models
that, for some reason, have a large $\theta_{13}^L$, or to a supersymmetric framework where the 
$C\!P$ asymmetry doubles and the wash-out at the production can occur in the three-flavoured regime
and be greatly reduced and in both these cases one can have the appearance of electron-dominated solutions.
The third reason is that in this way we can extend our analytic description and show agreement with 
the numerical results in the two examples of Fig.~2 and Fig.~3 also for the quantities in the electron
flavour ($K_{1e}$, $K_{2e}$, $\ve_{2e}$, $\eta_B^{(e)}$). This is not just an aesthetic reason but it provides
yet another cross check making us confident of the accuracy of our analytic solution. 

Analogously to $\ve_{2\mu}$ the electron $C\!P$ asymmetry can also be written as the sum of two terms,
\be
\ve_{2e} = \ve_{2e}^I + \ve_{2e}^{V_L} \,  .
\ee
The first one is the non-vanishing one in the limit for $V_L \ra I$ and is very strongly suppressed
\footnote{We are including it simply to describe $\ve_{2e}$ correctly even when $V_L \ra I$.}
 and given by
\be
\ve_{2e}^I =  {3 \,m^2_{D2} \over 16\, \pi \, v^2}\,{m_{D1}^4 \over m_{D2}^4}\, 
{m_{D2}^2 \over m_{D3}^2} \, 
 {|\widetilde{m}_{\nu 12}| \over m_1 \, m_2 \, m_3}\, 
 {|(\widetilde{m}_{\nu}^{-1})_{13}|\, |(\widetilde{m}_{\nu}^{-1})_{23}|\over |(\widetilde{m}_{\nu}^{-1})_{33}|^2}\,
 { |V_{L 11}|^2 \over |(\widetilde{m}_{\nu}^{-1})_{33}|^{2} + |(\widetilde{m}_{\nu}^{-1})_{23}|^{2}}
 \, \sin \widetilde{\a}^I_{Le} \,   ,
\ee
with 
\be
\widetilde{\a}^I_{Le} = {\rm Arg}\left[\widetilde{m}_{\nu 12} \right]  - {\rm Arg}[(\widetilde{m}^{-1}_{\nu})_{23}] - {\rm Arg}[(\widetilde{m}^{-1}_{\nu})_{13}]- \pi -2\,(\rho+\s) -2\,(\rho_L+\s_L)  \,  .
\ee
The second one dominates  when $V_{L} \simeq V_{CKM}$ and is given by
\be
\ve_{2e}^{V_L}  = 
{3 \,m^2_{D2} \over 16\, \pi \, v^2} \,
 {|\widetilde{m}_{\nu 11}| \over m_1 \, m_2 \, m_3}\, 
 {|(\widetilde{m}_{\nu}^{-1})_{23}|\over |(\widetilde{m}_{\nu}^{-1})_{33}|}\,
 { |V_{L 21}| \, |V_{L 31}| \over |(\widetilde{m}_{\nu}^{-1})_{33}|^{2} + |(\widetilde{m}_{\nu}^{-1})_{23}|^{2}}
 \, \sin \widetilde{\a}^I_{Le} \,  ,
\ee
with $\widetilde{\a}^I_{Le} = \a_{L\m}^A$. For the flavoured decay parameters the following
expressions accurately reproduce the numerical results
\be
K_{1e} \simeq {|\widetilde{m}_{\nu 11}| \over m_{\star}} \, 
\left(|V_{L11}|^2 -2\,|V_{L11}|\,|V_{L21}|\,{\rm Re}[\widetilde{m}_{\nu 12}/\widetilde{m}_{\nu 11}] 
+|V_{L21}|^2 \,|\widetilde{m}_{\nu 12}|^2/|\widetilde{m}_{\nu 11}|^2\right)  \,   ,
\ee
while for all purposes  $K_{2e}$ can be completely neglected in the wash-out at the production, thus 
entirely dominated by $K_{2\mu}$.
We conclude this subsection mentioning that in \cite{A2Zlep} electronic solutions had been
found including a term in the asymmetry generated by flavour coupling.  However these solutions require strong  fine-tuning in the seesaw formula  at the level of $0.1\%$. 

\section{Conclusions}

We obtained a full analytical description for the calculation of the baryon asymmetry in $SO(10)$-inspired leptogenesis, 
\footnote{The set of analytical expressions are summarised in the Appendix.} generalising the results obtained in \cite{SO10decription}
accounting for the misalignment between the Yukawa basis and the flavour basis described by a unitary matrix with mixing angles
at the level of the mixing angles in the CKM matrix in the quark sector. In this way we could provide an analytical insight  into $SO(10)$-inspired leptogenesis able to explain the relaxation of constraints, in particular the upper bound on the atmospheric
mixing angles in the case of strong thermal (tauon-dominated) solutions   and the appearance of muon-dominated solutions at large values of $m_1 \gtrsim m_{\rm sol}$. 
We have shown how the analytic solution we obtained does not just provide a qualitative understanding,
but in fact, within the given set of assumption, it reproduces accurately the  asymmetry calculated numerically and can be basically confidently used for the calculation of the baryon asymmetry in $SO(10)$-inspired leptogenesis without passing through the lengthy numerical diagonalisation of the Majorana mass matrix in the Yukawa basis. This solution provides a thorough analytic insight and paves the way for the account  of different effects in the derivation of the constraints on the low energy neutrino parameters,  including, importantly, their statistical significance, a crucial step in light of the expected future experimental progress, for the testability of $SO(10)$-inspired leptogenesis. 

\vspace{-1mm}
\subsection*{Acknowledgments}

PDB acknowledges financial support from the NExT/SEPnet Institute.   PDB acknowledges 
financial support from the STFC Consolidated Grant ST/L000296/1 . 
This project has received funding from the European Union's Horizon 2020 research and innovation programme under the Marie Sklodowska-Curie grant agreement No 690575 and No 674896. 
PDB is also grateful to the Tokyo University for its hospitality during the period this paper was prepared
and wishes to thank Koichi Hamaguchi for useful discussions. 

\newpage
\section*{Appendix}
\appendix
\renewcommand{\thesection}{\Alph{section}}
\renewcommand{\thesubsection}{\Alph{section}.\arabic{subsection}}
\def\theequation{\Alph{section}.\arabic{equation}}
\renewcommand{\thetable}{\arabic{table}}
\renewcommand{\thefigure}{\arabic{figure}}
\setcounter{section}{1}
\setcounter{equation}{0}

In this Appendix we summarise in a compact way all the set of analytical expressions that
constitute the solution of $SO(10)$-inspired leptogenesis we found. This set
can be easily plugged into a simple code for  a fast calculation of the asymmetry and
the generation of a big amount of solutions.  This is the set of needed equations:
\be
\widetilde{m}_{\nu} \equiv V_L\,m_{\nu}\,V_L^T  \,   ,
\ee

\bea
\Phi_1 & = & {\rm Arg}[-\widetilde{m}_{\nu 11}^{\star}] \,  ,  \\
\Phi_2 & = & {\rm Arg}\left[{\widetilde{m}_{\nu 11}\over (\widetilde{m}_{\nu}^{-1})_{33}}\right] -2\,(\rho+\s)-2\,(\rho_L + \s_L) \, ,  \\
\Phi_3 & = &  {\rm Arg}[-(\widetilde{m}_{\nu}^{-1})_{33}] \,  ,
\eea

\be
D_{\phi} \equiv {\rm diag}(e^{-i \, {\Phi_1 \over 2}}, e^{-i{\Phi_2 \over 2}}, e^{-i{\Phi_3 \over 2}}) \,  ,
\ee

\be
U_R \simeq  
\left( \begin{array}{ccc}
1 & -{m_{D1}\over m_{D2}} \,  {\widetilde{m}^\star_{\nu 1 2 }\over \widetilde{m}^\star_{\nu 11}}  & 
{m_{D1}\over m_{D3}}\,
{ (\widetilde{m}_{\n}^{-1})^{\star}_{13}\over (\widetilde{m}_{\n}^{-1})^{\star}_{33} }   \\
{m_{D1}\over m_{D2}} \,  {\widetilde{m}_{\nu 12}\over \widetilde{m}_{\nu 11}} & 1 & 
{m_{D2}\over m_{D3}}\, 
{(\widetilde{m}_{\n}^{-1})_{23}^{\star} \over (\widetilde{m}_{\n}^{-1})_{33}^{\star}}  \\
 {m_{D1}\over m_{D3}}\,{\widetilde{m}_{\nu 13}\over \widetilde{m}_{\nu 11}}  & 
- {m_{D2}\over m_{D3}}\, 
 {(\widetilde{m}_\nu^{-1})_{23}\over (\widetilde{m}_\nu^{-1})_{33}} 
  & 1 
\end{array}\right) 
\,  D_{\Phi} \,  ,
\ee

\bea
M_1   & \simeq  &   {\a_1^2 \, m^2_{\rm u} \over |(\widetilde{m}_\nu)_{11}|} \, , \\
M_2   &  \simeq &    {\a_2^2 \, m^2_{\rm c} \over m_1 \, m_2 \, m_3 } \, {|(\widetilde{m}_{\nu})_{11}| \over |(\widetilde{m}_{\nu}^{-1})_{33}|  } \,  ,  \\
M_3  & \simeq &   \a_3^2\, {m^2_{\rm t}}\,|(\widetilde{m}_{\nu}^{-1})_{33}|  \, ,
\eea

\be\label{KialAPPENDIX}
K_{i\a} = {\sum_{k,l} \, 
m_{Dk}\, m_{Dl} \,V_{L k\a} \, V_{L l \a}^{\star} \, U^{\star}_{R ki} \, U_{R l i} 
\over M_i \, m_{\star}}\,  ,
\ee

\be\label{ve2alAPPENDIX}
\ve_{2\a} \simeq {3 \over 16\, \pi \, v^2}\,
{|(\widetilde{m}_{\nu})_{11}| \over m_1 \, m_2 \, m_3}\,
{\sum_{k,l} \,m_{D k} \, m_{Dl}  \,  {\rm Im}[V_{L k \a }  \,  V^{\star}_{L  l \a } \, 
U^{\star}_{R k 2}\, U_{R l 3} \,U^{\star}_{R 3 2}\,U_{R 3 3}] 
\over |(\widetilde{m}_{\nu}^{-1})_{33}|^{2} + |(\widetilde{m}_{\nu}^{-1})_{23}|^{2}}   \,  ,
\ee

\bea
N_{\D_e}^{\rm lep,f} & \simeq &
\ve_{2e} \, \kappa(K_{2e} + K_{2\m}) \,
\, e^{-{3\pi\over 8}\,K_{1 e}}  \,   , \\ 
N_{\D_\m}^{\rm lep,f} & \simeq & 
\ve_{2\m}\,\kappa(K_{2e} + K_{2\m}) 
\, e^{-{3\pi\over 8}\,K_{1 \mu}} \,  , \\ 
N_{\D_\t}^{\rm lep,f} & \simeq & \ve_{2 \tau}\,\kappa(K_{2 \tau})\,e^{-{3\pi\over 8}\,K_{1 \tau}} \,  ,
\eea

\be
N_{B-L}^{\rm p, f} = \sum_{\a} \,  N_{\D_\a}^{\rm p,f}   \,   ,
\ee
and finally
\be
\eta_B^{\rm lep} =a_{\rm sph}\,{N_{B-L}^{\rm lep,f}\over N_{\g}^{\rm rec}} \simeq 
0.96\times 10^{-2}\,N_{B-L}^{\rm lep,f} \,  .
\ee
Notice that all these expressions are valid for any $V_L$, it indeed relies only on the first assumption
of $SO(10)$-inspired leptogenesis (hierarchical Yukawas) and not on the second, small angles in $V_L$.
This can be checked easily simply taking as an example the extreme case when all leptonic mixing comes from $V_L$, in a way that $V_L = U^{\dagger}$ and $U_R = I$. In this case simply $\widetilde{m}_\nu = - D_m$ and
$M_1 = m^2_{D1}/m_1$, $M_2= m^2_{D2}/m_2$ and $M_3 = m^2_{D3}/m_3$, as it has to be
considering that $\Omega = I$. One can indeed also check that the analytic expression for the $\O$ matrix Eq.~(\ref{Omegaapp}) reduces to $\Omega = I$.  
This shows that the analytical expressions are consistent with a choice of $V_L$
that is very different from $V_L \simeq V_{CKM}$. Notice, however, that in this case it is not guaranteed that 
$M_1 \lesssim 10^{9}\,{\rm GeV}$ and so that the $N_2$-dominated scenario of leptogenesis holds. For this 
to be true, one needs also to impose $I \leq V_L \lesssim V_{CKM}$. For small mixing angles $\theta_{ij}^L$ one can extract the leading terms in the sums in Eqs.~(\ref{KialAPPENDIX}) and (\ref{ve2alAPPENDIX}) obtaining the explicit analytic expressions we showed in the body text and that we do not repeat here.

\end{document}